# *Nanoinsulators and nanoconnectors for optical nanocircuits*

*Mário G. Silveirinha*[(1,2)], *Andrea Alù*[(1)], *Jingjing Li*[(1)], *and Nader Engheta*[(1,\*)]

(1) University of Pennsylvania, Department of Electrical and Systems Engineering, Philadelphia, PA, U.S.A., <u>engheta@ee.upenn.edu</u>

(2) Universidade de Coimbra, Department of Electrical Engineering – Instituto de Telecomunicações, Portugal

**Abstract**

Following our recent idea of using plasmonic and non-plasmonic nanoparticles as nanoinductors and nanocapacitors in the infrared and optical domains [N. Engheta, A. Salandrino, and A. Alù, *Phys. Rev. Letts.*, Vol. 95, 095504, (2005)], in this work we analyze in detail some complex circuit configurations involving series and parallel combinations of these lumped nanocircuit elements at optical frequencies. Using numerical simulations, it is demonstrated that, after a proper design, the behavior of these nanoelements may closely mimic that of their lower frequency (i.e., radio frequency (RF) and microwave) counterparts, even in relatively complex configurations. In addition, we analyze here in detail the concepts of nanoinsulators and nanoconnectors in the optical domain, demonstrating how these components may be crucial in minimizing the coupling between adjacent optical nanocircuit elements and in properly connecting different branches of the nanocircuit. The unit nanomodules for lumped nanoelements are introduced as building blocks for more complex nanocircuits at optical frequencies. Numerical simulations of some complex circuit scenarios considering the frequency response of these nanocircuits are presented and discussed in details, showing how practical applications of such optical nanocircuit concepts may indeed be feasible within the current limits of nanotechnology.

PACS numbers: 61.46.-w, 07.50.Ek, 78.20.-e, 78.67.Bf

---

[*] To whom correspondence should be addressed: E-mail: engheta@ee.upenn.edu

# I. Introduction

The interdisciplinary field of nanotechnology is today one of the most important and exciting research areas in science. The interaction of optical waves with nanoparticles is currently one of the important problems in this field. In a recent work [1], we have suggested that since the size of nanoparticles may be much smaller than the wavelength of optical waves, they may be treated as "lumped nanocircuit elements". This concept is very appealing because it may allow envisioning an extension of standard low frequency modular circuit technology to the infrared and optical domains, with all the implications that this would have in a wide range of applications. As pointed out in [1], a mere scaling of the circuit components used at radio and lower frequencies to the infrared and optical domains may not work, because metals change their conducting properties in the optical domain [2]. Instead, in [1] we have suggested to use arrangements of plasmonic and non-plasmonic particles to design complex optical nanocircuits, and in particular we have demonstrated that plasmonic and non-plasmonic nanoparticles may effectively act as nanoinductors and nanocapacitors, respectively. We have envisioned several coupled nanoscale circuit configurations that are the analogues of the standard low frequency parallel and series combinations of lumped elements. Moreover, at optical frequencies we have considered the displacement current $-i\omega\mathbf{D}$ (with $\omega$ being the radian frequency of operation and $\mathbf{D}$ the local electric displacement vector inside the nanoparticles) as the counterpart of the electric current density $\mathbf{J}_c$ in conductors at low frequencies. Using these ideas and analogies, we have proposed a new design for the optical implementation of right-handed and left-handed planar nanotransmission lines, and in particular in [3] we have shown how such transmission lines may be synthesized using layered plasmonic and non-plasmonic materials and how in many ways their characteristics are similar to those of their lower-frequency transmission-line counterparts. In [4] we have also applied these concepts to linear cascades of plasmonic and non-plasmonic nanoparticles, showing how they may mimic the regular cascades of inductors and capacitors at lower frequencies in order to realize nanowires and nanotransmission lines. In [5], moreover, we have extended these concepts to 3D arrangements of nanoparticles to envision complex 3D nanocircuit and nanotransmission line metamaterials with anomalous properties and an effective negative index of refraction. In [6] we presented the results of our analysis on parallel and series combinations of nanoelements and some anomalous properties arising in simple resonant configurations, again analogous to their low-frequency counterparts. Finally, in [7], we presented our model of the coupling among neighboring nanocircuit elements, showing the main limitations and complications that the simple approach of placing lumped nanocircuit elements in the close vicinity of each other may have. Recently, in [8] a method based on electronic structure (nonclassical) theory was used to determine the equivalent circuit representations of nanostructured physical systems at optical frequencies.

Despite these recent development, the design of optical nanocircuits may still pose some theoretical and, of course, technological challenges. An important aspect discussed in details in the following is that, unlike its low frequency equivalents, the proposed optical nanocapacitors, nanoinductors and nanoresistors may suffer from displacement current leakage which may adversely affect the overall performance of the system. Indeed, while at low frequencies the electric current density $\mathbf{J}_c$ is confined to the conductor surface because the background materials (i.e., free space) have a very poor or zero conductivity, its optical circuit counterpart $-i\omega\mathbf{D}$ may in general leak out of the branches of the circuit, interacting with the surrounding region and establishing a strong coupling among the different lumped nanoelements. Another important problem is represented by the optical interconnection between lumped nanocircuit elements not necessarily adjacent to each other. It is demonstrated here that due to the strong geometrical and polaritonic resonances of the materials near the junctions of "lumped" elements, the behavior of a straightforward realization of the proposed nanocircuits may be different from what is desired in many ways.

To circumvent these problems, in this work we analyze in detail the concepts of optical *nanoinsulators* and *nanoconnectors* in optical nanocircuits. We demonstrate that the displacement current leakage from nanocircuit elements may be avoided by properly covering these nanoelements with a suitable "shield" made of a material with permittivity $\varepsilon$-near zero (ENZ). Such materials may be readily available at infrared and optical frequencies where some low-loss metals (Au, Ag) [9]-[11], some semiconductors [12], and polar dielectrics such as Silicon Carbide (SiC) [13] already possess permittivity near zero. Otherwise they may in principle be constructed by nanostructuring available materials using metamaterial theory [14, 15, 16]. Interestingly, in recent works it was demonstrated that materials with permittivity near zero may play interesting roles in seemingly unrelated problems such as: to transform curved wavefronts into planar ones and to design delay lines [17]-[18], to narrow the far-field pattern of an antenna embedded in the medium [19] or to induce anomalous cloaking phenomena [20]. Also, recently we demonstrated that ENZ materials may be used to squeeze electromagnetic waves through subwavelength channels and waveguides [21]. In this work, we aim at using such materials as optical nanoinsulators for the displacement current in lumped nanocircuits. Indeed, as detailed ahead, our theoretical analysis shows that at optical wavelengths such layers may, under certain circumstances, act as insulators supporting zero displacement current, resulting in the confinement of the displacement current inside the optical nanoparticles. In analogy with materials with very low conductivity in the classical circuit concepts, here low permittivity materials may play an analogous role for the displacement current in the optical domain. Therefore, ENZ-shielded nanocircuit elements may indeed be regarded intuitively as lumped elements with lower leakage coupling among neighboring nanoelements. On the other hand, we will show that (plasmonic or nonplasmonic) materials with relatively large permittivity may be used as optical nanoconnectors, and may effectively interconnect different lumped nanoelements without inducing strong geometrical or polaritonic resonances. In this work, we will present our recent theoretical and numerical findings in these matters, and we will forecast some future ideas and potential applications of these concepts.

This paper is organized as follows. In section II, we generalize the concepts introduced in [1], and derive simple circuit models for optical nanocircuits with nanowires as building blocks. Using a simple computational model, we study the performance of straightforward realizations of the envisioned optical nanocircuits. In section III, the optical nanoinsulator concept is introduced, and it is shown that it may be possible to force the induced displacement current to flow within the nanocircuit boundaries by properly shielding the proposed nanoelements with ENZ materials. We characterize series and parallel arrangements of the insulated nanoelements in relevant scenarios, and compare their behavior with the proposed circuit models. In section IV, we demonstrate that it is possible to improve the "connection" between the nanoelements, eradicating possible geometrical and polaritonic resonances at the junctions, by using optical nanoconnectors made of materials with relatively large permittivity. In sections V and VI we apply these concepts to more complex 3D scenarios, in order to envision realizable nanocircuits relying on parallel or series interconnections, and we verify numerically our intuitions in these more complex configurations. Finally, in section VII the conclusions are drawn. In this manuscript we assume that the electromagnetic fields have the time variation $e^{-i\omega t}$.

## II. Nanocircuit analogy

In this section we briefly review the concepts and ideas originally introduced in [1], and we test numerically the performance of straightforward realizations of these optical lumped nanocircuits using a full wave electromagnetic simulator.

In [1] it was shown that the interaction of an impressed field with a sub-wavelength spherical particle standing in free-space may be conveniently described using circuit theory concepts. The equivalent circuit model for the spherical particle is either a nanocapacitor or a nanoinductor, depending on the real part of the permittivity of the nanosphere being positive, $\text{Re}\{\varepsilon\} > 0$, or negative, $\text{Re}\{\varepsilon\} < 0$, respectively. In addition, the imaginary part of the material permittivity may provide an equivalent nanoresistor. Our objective here is to analyze in detail the electromagnetic behavior of such nanoparticles when arranged in a series or parallel circuit configuration. Although the spherical geometry may be appealing from a mathematical point of view for its simplicity, it is less appropriate for configurations in which one wants to physically "connect" many of these nanoparticles; indeed, two non-overlapping spheres can at most intersect in one point, and therefore it may be difficult to connect them in a complex nanocircuit platform without generating undesired coupling phenomena. For this reason we have analytically solved the problem of two conjoined half-cylinders, as presented in [6], which indeed may look like two nanocircuit elements connected in parallel or series, depending on the orientation of the applied electric field with respect to their common interface. In the present work, more in general we assume that our particles are shaped as sub-wavelength nanocylinders or nanowires, as illustrated in Fig. 1. For simplicity, in our mathematical model we admit that the nanowires have uniform cross-section $A_T$ and may have a certain radius of curvature.

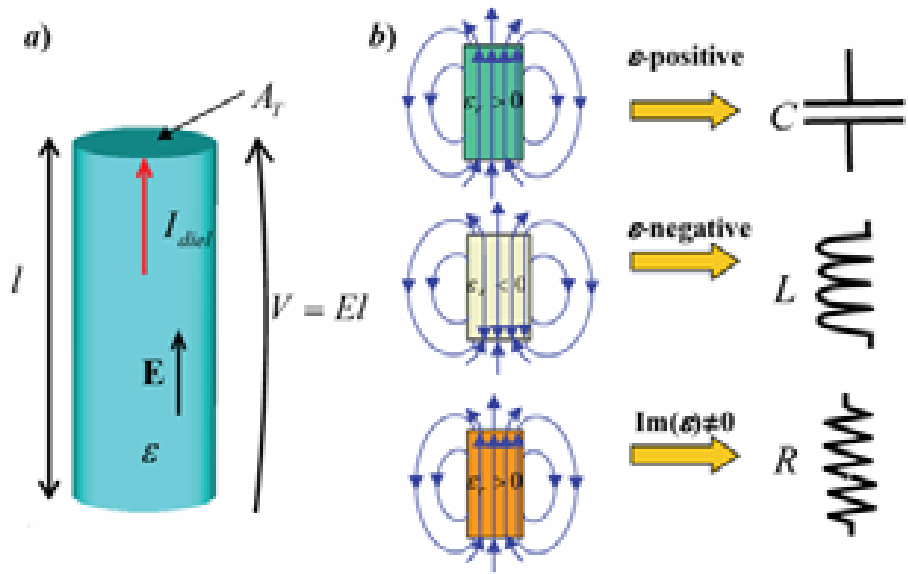

**Fig. 1.** (Color online) Panel *a*: Geometry of a generic subwavelength nanocircuit element in the form of a nanowire with length *l* and cross-section $A_T$. Panel *b*: equivalent circuit model for the nanowire depending on the electrical properties of the material. The sketch of electric field lines inside the nanowire are also shown (blue - dark in grayscale - arrows).

To begin with, let us analyze the electromagnetic properties of such subwavelength wires, namely their equivalent circuit impedance. To this end, let $E$ be the electric field inside the subwavelength particle (see Fig. 1). As proposed in [1], we can regard the displacement current $-i\omega \mathbf{D}$ as the current density flowing inside the nanowire, where $\mathbf{D} = \varepsilon \mathbf{E}$ is the electric displacement vector. Thus, the total current flowing through the cross-section of the nanowire is given by:

$$I_{diel} = -i\omega \phi_e, \qquad \phi_e = \int_{\text{cross-section}} \mathbf{D} \cdot \mathbf{ds} = DA_T = \varepsilon E A_T \qquad (1)$$

where $\phi_e$ is by definition the electric flux through the cross-section. In general, the current $I_{diel}$ is not uniform along the wire axis. The reason for this phenomenon is that the electric field lines are not confined to the nanoelement, and so part of the current continuously leaks out through the lateral walls into the background region. Indeed, from Gauss's law $\nabla \cdot \mathbf{D} = 0$, one can easily find that $\phi_{e,2} - \phi_{e,1} = \phi_{leak}$, where $\phi_{e,1}$ and $\phi_{e,2}$ are the electric fluxes through the cross-section of the nanowire in two distinct cuts along its axis, and $\phi_{leak}$ is the electric flux through the walls delimited by the referred transverse cuts. Thus, the equivalent current $I_{diel} = -i\omega \phi_e$ may be uniform along the nanoelement if and only if the leakage through the lateral walls is negligible, i.e., $\phi_{leak} = 0$. Note that this effect was recognized in our previous work [1], and properly taken into account by modeling the free-space region as an equivalent fringe capacitance in parallel with the equivalent impedance of the subwavelength particle. It is also important to point out that the leakage of the displacement current through the lateral walls of the nanowire is to some extent a phenomenon very specific of the proposed circuit configuration, and has no direct analogue at low frequencies. In fact, in regular conductors the electric conduction current is naturally confined to a region close to the surface of the material since it involves the drift of free electric charges and in general the background material has poor, if not zero, conductivity. Quite differently, in our optical nanocircuits the equivalent displacement currents are mostly associated to oscillations of electric dipoles induced in the material (at least for regular dielectrics), and not specifically to the drift of free-charges, and therefore a non-zero permittivity in the background material would be sufficient to induce an equivalent current leakage.

Let us assume temporarily that the leakage flux is approximately zero, $\phi_{leak} \approx 0$. In this case, as referred above, the current $I_{diel}$ and the flux $\phi_e$ are uniform inside and along the length of the nanoparticle. Since we assume that the cross-section and permittivity $\varepsilon$ of the subwavelength wire are uniform, it is clear that the electric field is also necessarily uniform, and thus the voltage drop across the length $l$ is given by $V = \int \mathbf{E} \cdot \mathbf{dl} = El$ (see Fig. 1). Consequently, the equivalent impedance of the nanocircuit element is,

$$\bar{Z} \equiv \frac{V}{I_{diel}} = \frac{1}{-i\omega} \frac{V}{\phi_e} = \frac{1}{-i\omega} \Re_e, \qquad \Re_e = \frac{l}{\varepsilon A_T} \qquad (2)$$

where $\Re_e$ is by definition the *electric reluctance* of the material [F$^{-1}$]. The motivation for this designation is the parallelism that may be made between the theory developed here and the classical theory of magnetic circuits used to characterize transformers and other magnetic systems [22]. In fact, it may be verified that the problem under study is to some extent the electromagnetic dual of the classical problem of magnetic circuits. From (1) and (2) one also obtains the relation:

$$V = \Re_e \phi_e = \bar{Z} I_{diel} \qquad (3)$$

We stress that the derived results are valid only if the leakage through the lateral walls is zero. Also it is obvious that the formulas remain valid even when the nanocircuit element has a more complex shape with non-uniform radius of curvature (the only restriction is that its cross-section and permittivity remain constant). Equation (2) shows that when the permittivity of the nanoelement is positive $\varepsilon > 0$ (e.g., regular dielectric) its impedance is positive imaginary, whereas if the (real part of the) permittivity of the nanoelement is negative $\varepsilon < 0$ (e.g., plasmonic material) the impedance is negative imaginary. Consequently, in the lossless limit $\text{Im}(\varepsilon) \approx 0$, it is clear that a nanoelement with $\varepsilon > 0$ is equivalent to a nanocapacitor *C*, and a nanoelement with $\varepsilon < 0$ is equivalent to a nanoinductor *L*, given by:

$$C = \frac{1}{\Re_e} = \frac{\varepsilon A_T}{l}, \qquad L = -\frac{\Re_e}{\omega^2} = \frac{1}{\omega^2} \frac{l}{|\varepsilon| A_T} \qquad (4)$$

The referred circuit equivalence is schematized in panel *b* of Fig. 1. This result is completely consistent with our previous work [1] where the analysis was focused on spherical nanoparticles for simplicity. It is worth noting that (4) shows how it is possible to adjust the values of the equivalent *C* and *L* by properly selecting the size, shape, and material contents of the nanostructure, as pointed out in [1]. In the case of losses, i.e., $\text{Im}(\varepsilon) \neq 0$, the impedance $\bar{Z}$ has a real component that represents the effect of dissipation in the material. In that case, the equivalent model for the nanoparticles consists of a capacitive or inductive element in parallel or series (respectively) with a nanoresistor. When the losses are dominant, the subwavelength nanowire may be modeled using just an equivalent nanoresistor (see panel *b* of Fig. 1).

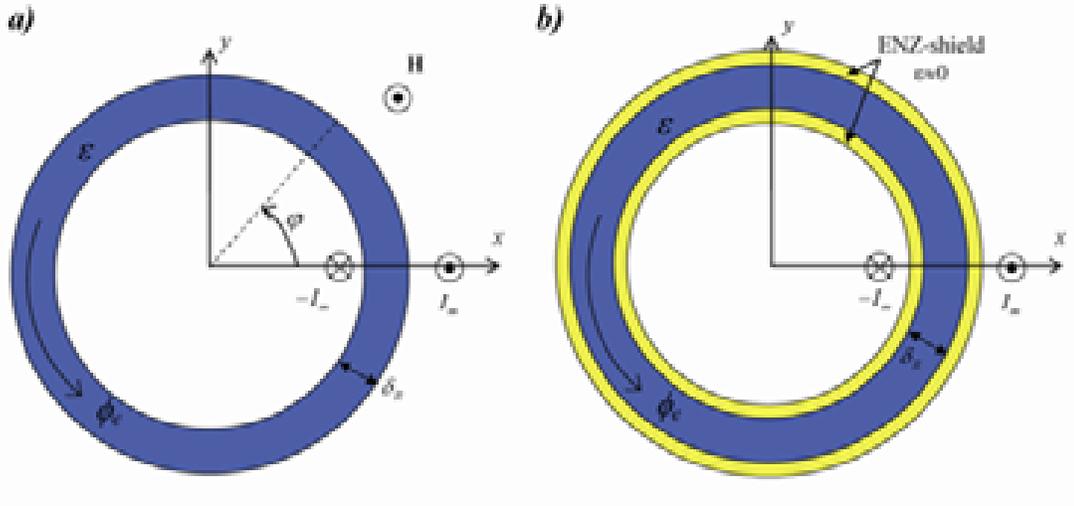

**Fig. 2.** (Color online) Geometry of a nanocircuit in the form of a ring with permittivity $\varepsilon$ fed by a balanced pair of fictitious magnetic line currents that induce an electromotive force across the circuit. Panel *a*: unshielded ring. Panel *b*: the ring is covered with an ENZ shield.

In order to understand the merits and limitations of this elementary model, we have performed several numerical experiments to test relevant arrangements and configurations of the proposed nanocircuit elements. For simplicity, we consider for now that the geometry is two-dimensional (2D), being the structure uniform along the *z*-direction and the magnetic field is such that $\mathbf{H} = H_z \hat{\mathbf{u}}_z$. Also, we will temporarily assume that the material loss is negligible. The geometry of the first scenario is depicted in panel *a* of Fig. 2. It consists of a ring with permittivity $\varepsilon$ delimited by the region $R_1 < r < R_2$, where $(r, \varphi)$ is a system of polar coordinates defined with respect to the center of the ring. In the following simulations we will complicate this structure by adding different nanocircuit elements around the ring to simulate parallel and series interconnections in a closed-loop circuit. In this way we can simulate a basic "closed nano-circuit", which may help in understanding the coupling issues in a small and simple circuit network, analogous to a conventional lower-frequency circuit. The first important issue is how to properly excite the ring, and induce an electromotive force across the flux path. In [1] we have suggested to excite the nanocircuit using a local electric field, e.g., by using a near-field scanning optical microscope (NSOM). Here, to ease the numerical simulation we use a completely different feed, exploring the previously referred analogy between our problem and the theory of magnetic systems. Indeed, we note that a standard magnetic circuit (e.g. a transformer) is usually fed by encircling a coil of electric current around the magnetic core of the circuit. The electromagnetic dual of this configuration consists of a dielectric ring fed by a magnetic current wrapped around the core.

Using this analogy, we propose to feed our subwavelength ring with a fictitious pair of magnetic line sources with symmetric amplitude $I_m$ (Fig. 2). The magnetic line sources are placed along the *x*-axis at the positions $x = R_s^+$ and $x = R_s^-$. Since the density of magnetic current $\mathbf{J}_m$ is different from zero, Faraday's law becomes $\nabla \times \mathbf{E} = +i\omega\mathbf{B} - \mathbf{J}_m$. In the quasi-static limit the term $+i\omega\mathbf{B}$ may be neglected, and so the electromotive force across a closed path delimited by the line sources (oriented counterclockwise) is $V = NI_m$, where $N$ is the number of the turns of our equivalent coil (*N*=1 in Fig. 2). It can be easily verified that the magnetic field (directed along *z*) radiated by the line source located at $x = R_s^+$ is given by $H_z^{inc,+} = i\omega\varepsilon_0 I_m \Phi_0$ where $\Phi_0$ is the (free-space) Green function for a 2D-problem: $\Phi_0 = \dfrac{1}{-4i} H_0^{(1)}\left(k_0 |\mathbf{r} - \mathbf{r}'|\right)$ ($k_0 = \omega\sqrt{\varepsilon_0\mu_0}$ is the free-space wave number, and $H_n^{(1)} = J_n + iY_n$ is the Hankel function of 1$^{st}$ kind and order *n* [23]). In our simulation we have chosen $I_m$ such that the induced electromotive force is $V = 1$ [V]. We underline here that this feeding mechanism was chosen only to ease the numerical simulations and test the validity of our circuit models. Ahead in the paper, we will present results for a more realistic model of the feed at optical wavelength.

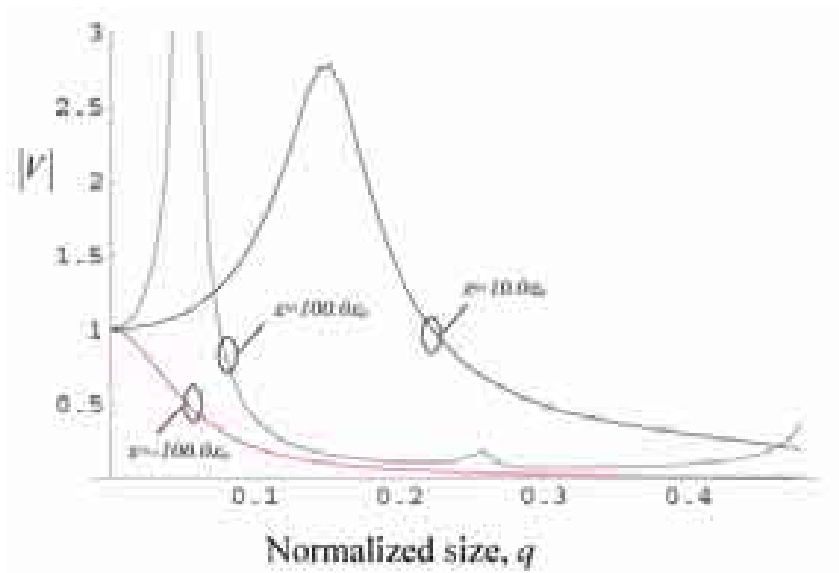

**Fig. 3.** (Color online) Amplitude of the induced voltage [in Volt] along the path $r = R_{med}$, for different values of the permittivity $\varepsilon$ of the subwavelength ring.

In our simulations, the dimensions of the ring were set equal to $R_1 = 0.8q\lambda_0$ and $R_2 = 1.0q\lambda_0$, and the line sources were positioned at $R_s^+ = 1.1q\lambda_0$ and $R_s^- = 0.7q\lambda_0$, where $\lambda_0$ is the free-space wavelength, and $q$ is some (dimensionless) quantity that defines the electrical size of the structure. In order to check the validity of the quasi-static approximation $\nabla \times \mathbf{E} \approx -\mathbf{J}_m$, i.e., if the term $+i\omega\mathbf{B}$ is negligible when compared to $-\mathbf{J}_m$, we have computed numerically the electromotive force $V$ as a function of $q$ along the path $r = R_{med}$, with $R_{med} = 0.5(R_1 + R_2)$. To this end, the Maxwell equations have been solved numerically using a dedicated full wave numerical code that implements the method of moments (MoM). The result is reported in Fig. 3 for different values of the permittivity of the ring. It is seen that for $q < 0.02$ the induced voltage is approximately 1 [V] for all the considered examples, and consequently only under this condition the quasi-static approximation is valid. All the results presented in the following of this section are computed assuming $q = 0.002$.

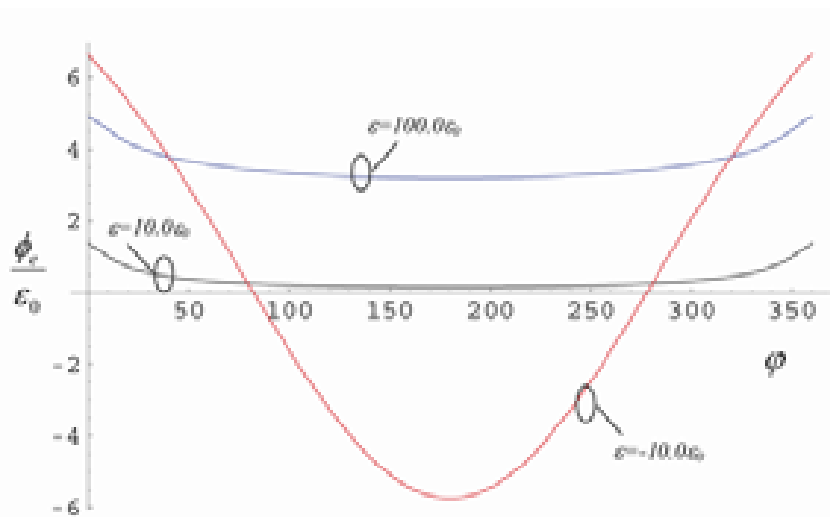

**Fig. 4.** (Color online) Normalized electric flux (p.u.l.) inside the subwavelength ring as a function of $\varphi$, for different values of the ring permittivity.

In order to evaluate the relative importance of the leakage flux $\phi_{leak}$ through the lateral walls, we have computed numerically the flux $\phi_e$ inside the subwavelength ring (see Fig. 2). Note that since the problem under study is two-dimensional and the structure is uniform along the $z$-direction, it is meaningful to compute the flux per unit length (p.u.l). To keep the notation simple, we also represent the flux p.u.l with the symbol $\phi_e$. To a first approximation, we can write $\phi_e \approx \varepsilon E_\varphi \delta_R$, where $\delta_R = R_2 - R_1$ defines the cross-section of the ring. Similarly, the inverse of the electric reluctance $1/\Re_e$ is also specified in p.u.l unities. For the subwavelength ring shown in Fig. 2 the 2D-reluctance is given by

$$\Re_e = \frac{2\pi R_{med}}{\varepsilon \delta_R}$$ (compare with (2)).

The computed flux $\phi_e$ is shown in Fig. 4 as a function of the angle $\varphi$ and for different values of the permittivity of the ring material. Note that in reality $\phi_e$ is a complex number, but since the dimensions of the ring are very small as compared to the wavelength of radiation, the imaginary part of $\phi_e$ is always negligible. Very disappointingly, it is seen that $\phi_e$ may depend relatively strongly on $\varphi$ (particularly near the two line sources, i.e., at $\varphi = 0$), and consequently it cannot be considered uniform inside the ring. This evidently demonstrates that in general the leakage flux is not negligible, and that therefore the subwavelength ring may have a strong coupling with the neighboring free-space region. This is particularly true in the case $\varepsilon = -10\varepsilon_0$ where the induced flux varies noticeably inside the ring. The only case in which the flux is nearly uniform is when $\varepsilon = 100\varepsilon_0$, i.e., for relatively large positive values of the ring permittivity. Since leakage flux is not negligible, we cannot apply directly (2) and (3), and our simplified circuit model is not adequate for this case. Indeed, it is clear that an additional fringe capacitance should be considered here in order to properly model the coupling of this nanowire with the free-space region, as proposed in [1] and further presented in [7], but this may complicate the design of a complex nanocircuit system, when/if the coupling among lumped nanoelements is undesirable. Nevertheless, it can be verified that the modified relation $V = \Re_e \langle \phi_e \rangle$ holds, where $\langle \phi_e \rangle$ is average flux across the flux path. For example, in the case $\varepsilon = -10\varepsilon_0$ our numerical calculations (obtained by averaging $\phi_e$ depicted in Fig. 4 over $\varphi$) show that $\langle \phi_e \rangle / \varepsilon_0 = -0.353$ [V]. On the other hand, the theoretical value of the reluctance is $\Re_e = -0.353^{-1}/\varepsilon_0 = -2.83/\varepsilon_0$ [F/m]$^{-1}$ (which corresponds to a nanoinductor in the circuit model). Both values are consistent with fact that the magnetic line sources induce an electromotive force $V = 1$ [V] along the flux path and with Eq. (3).

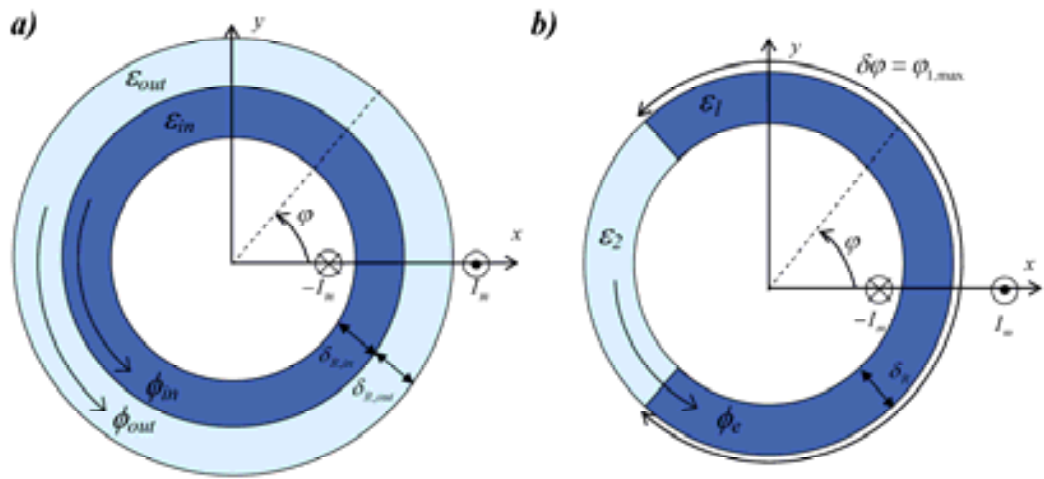

**Fig. 5.** (Color online) Panel *a*: Geometry of two concentric subwavelength rings arranged in a parallel circuit configuration. Panel *b*: Geometry of two subwavelength ring sections arranged in a series circuit configuration. In both cases the equivalent circuit is fed by a balanced pair of magnetic line sources.

To further illustrate the problems related with the flux leakage and the strong coupling between neighboring circuit components, let us consider the configuration depicted in panel *a* of Fig. 5. It shows two concentric rings defined by $R_1 < r < R_{int}$ (inner ring) and $R_{int} < r < R_2$ (outer ring), where $r = R_{int}$ defines the interface between the two rings. The thickness of the inner/outer ring is $\delta_{R,in} = R_{int} - R_1$ and $\delta_{R,out} = R_2 - R_{int}$, respectively. In the simulations we considered that $R_1 = 0.8q\lambda_0$, $R_{int} = 0.9q\lambda_0$, and $R_2 = 1.0q\lambda_0$, with $q = 0.002$. The permittivity of the rings is $\varepsilon_{in}$ (inner ring) and $\varepsilon_{out}$ (outer ring). The rings are fed with the same line source configuration as in the previous example. Based on the ideas presented in [1], one may expect that the equivalent circuit for this ring arrangement consists of the parallel combination of the impedances of the individual rings fed by the equivalent voltage generator or, in other words, that the equivalent impedance of the system is the parallel combination of the individual impedances. In fact, it is clear that if the leakage flux through the walls $r = R_1$, $r = R_{int}$, and $r = R_2$ is negligible, then the flux inside the inner ring, $\phi_{in}$, and the flux inside the outer ring, $\phi_{out}$, must be nearly constant (see Fig. 5). Since the voltage drop along the flux path is $V$ in both cases, one concludes that:

$$V = \Re_{in}\phi_{in} \quad ; \quad V = \Re_{out}\phi_{out} \tag{5a}$$
$$\phi_{e,tot} = \phi_{in} + \phi_{out} \tag{5b}$$

where $\Re_{e,in} = \dfrac{2\pi R_{med,in}}{\varepsilon_{in}\delta_{R,in}}$ is the 2D-reluctance of the inner ring, and $\Re_{out}$ is defined similarly. Note that $\phi_{e,tot}$ defined as above is the total flux (p.u.l) through the cross-section combined system. The equivalent reluctance of the system is given by $\Re_{eq} \equiv \dfrac{V}{\phi_{e,tot}}$. From the above relations it is immediate that,

$$\frac{1}{\Re_{eq}} = \frac{1}{\Re_{in}} + \frac{1}{\Re_{out}} \tag{6}$$

i.e., as we have anticipated, the equivalent circuit model is the parallel combination of the individual nanocircuit elements. Note the above result is exact in the quasi-static limit, and only assumes that the leakage flux is negligible.

To test these hypotheses and the proposed model, we have computed numerically the fluxes inside the two rings for several values of the permittivities. In Fig. 6, the normalized $\phi_{in}$ and $\phi_{out}$ are depicted as a function of the azimuthal angle, for the case $\varepsilon_{in} = 20.0\varepsilon_0$ and $\varepsilon_{out} = 10.0\varepsilon_0$ (solid lines). This corresponds to the parallel combination of two nanocapacitors. As in the previous example, the fluxes vary appreciably inside the ring, particularly near the line sources ($\varphi = 0$). Also, it may be seen that the two rings are not completely uncoupled, because flux $\phi_{in}$ ($\phi_{out}$) is slightly perturbed when the outer (inner ring) is removed from the system (dashed lines).

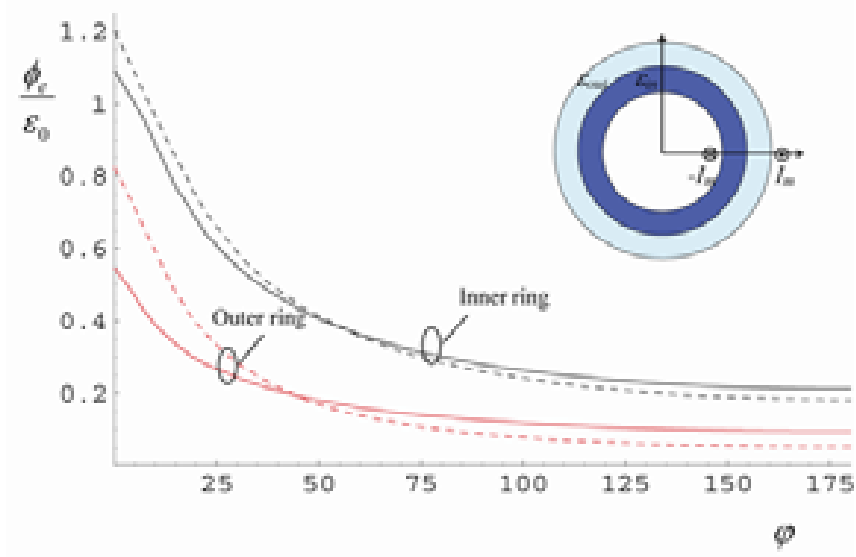

**Fig. 6.** (Color online) Normalized electric flux (p.u.l.) (solid lines) inside the subwavelength rings as a function of $\varphi$ for $\varepsilon_{in} = 20.0\varepsilon_0$ and $\varepsilon_{out} = 10.0\varepsilon_0$. The inset shows the geometry of the system. The dashed lines show the flux when only one ring is present and the other one is removed.

These results demonstrate that the leakage flux through the three interfaces is not negligible. In particular, our model (5)-(6) and the straightforward circuit analogy may not be completely and straightforwardly applied in this case. Nevertheless, it may be verified that, as in the previous example, the modified equations $V = \mathfrak{R}_{out} \langle \phi_{in} \rangle$ and $V = \mathfrak{R}_{out} \langle \phi_{out} \rangle$ are still accurate.

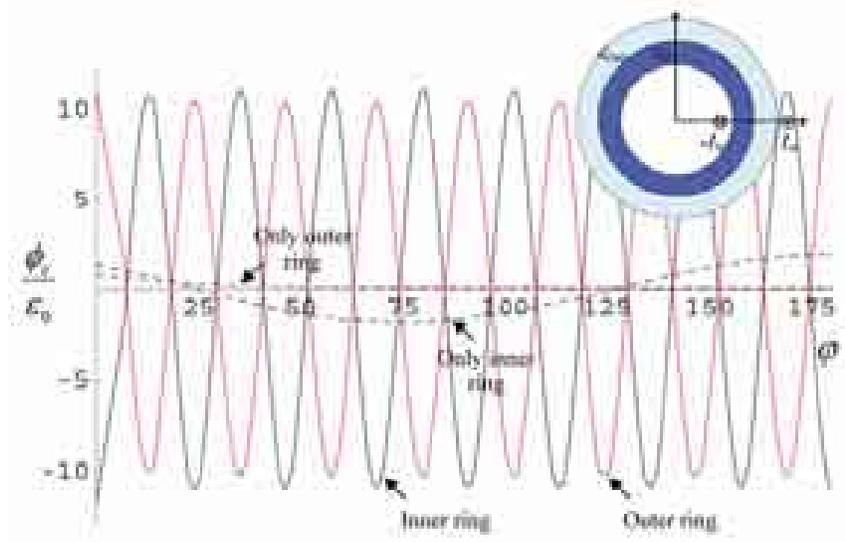

**Fig. 7.** (Color online) Similar to Fig. 6, but with $\varepsilon_{in} = -10.0\varepsilon_0$ and $\varepsilon_{out} = 10.0\varepsilon_0$.

The coupling between the two rings may be prominent when the respective permittivities have opposite signs and similar absolute values; in terms of a circuit model this case corresponds to the parallel association of a nanoinductor and a nanocapacitor. This effect is illustrated in Fig. 7 for $\varepsilon_{in} = -10.0\varepsilon_0$ and $\varepsilon_{out} = 10.0\varepsilon_0$ (solid lines). It is seen that the flux becomes highly oscillatory inside the rings, which suggest the excitation of a resonance, consistent with the circuit model. The dashed lines of Fig. 7 show that if one of the rings is removed the oscillatory response disappears. This result clearly shows that the observed resonance emerges due to the strong coupling between the two rings. This resonance is indeed closely related to the excitation of surface plasmon polaritons at the interface between a plasmonic and a non-plasmonic material.

As a final example, we report a configuration in which the nanoparticles are "connected" in series. The geometry is shown in panel *b* of Fig. 5. It consists of two ring sections juxtaposed in series, following the ideas and analogies proposed in [1]. The rings are delimited by the region $R_1 < r < R_2$ ($R_1$ and $R_2$ are chosen as in the previous examples), and are fed by the same source configuration as in the previous simulations. The ring with permittivity $\varepsilon_1$ fills the angular sector $|\varphi| < \varphi_{1,\max}/2$, and the ring with permittivity $\varepsilon_2$ fills the complementary region. In case the flux leakage through the walls $r = R_1$ and $r = R_2$ is negligible, it is obvious from Gauss's law $\nabla \cdot \mathbf{D} = 0$ that the flux $\phi_e$ is uniform inside the two rings and equal in both sections. In that case, it is easy to prove that,

$$V_1 = \Re_1 \phi_e \quad ; \quad V_2 = \Re_2 \phi_e \tag{7a}$$
$$V = V_1 + V_2 \tag{7b}$$

where $V_1$ and $V_2$ are respectively the (counterclockwise) voltage drops along ring-1 and ring-2, and $\Re_1 = \dfrac{\varphi_{1,\max} R_{med}}{\varepsilon_1 \delta_R}$ and $\Re_2 = \dfrac{(2\pi - \varphi_{1,\max}) R_{med}}{\varepsilon_2 \delta_R}$. Hence, the equivalent reluctance $\Re_{eq} \equiv \dfrac{V}{\phi_e}$ verifies,

$$\Re_{eq} = \Re_1 + \Re_2 \tag{8}$$

i.e., it is the series combination of the individual nanocircuit elements. However, as in the previous examples, this simplistic model may be of limited use, because the flux leakage may be a preponderant phenomenon. This is illustrated in Fig. 8 for different values of $(\varepsilon_1, \varepsilon_2, \varphi_{1,\max})$, where it is seen that the flux may appreciably vary with the azimuthal angle, particularly when $\varepsilon_1 = -20.0\varepsilon_0$, $\varepsilon_2 = 10.0\varepsilon_0$, $\varphi_{1,\max} = 180[\text{deg}]$. Notice that for this specific set of parameters the equivalent circuit model is the series association of a nanoinductor and a nanocapacitor.

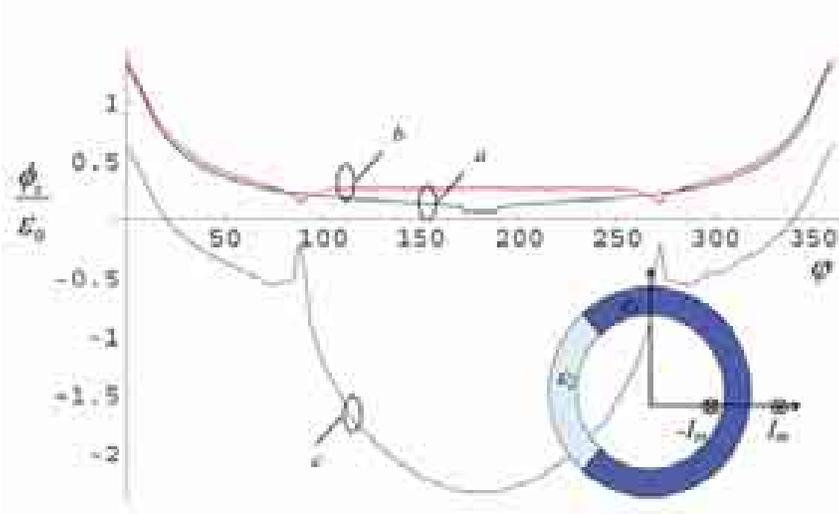

**Fig. 8.** (Color online) Normalized electric flux (p.u.l.) inside the subwavelength ring as a function of $\varphi$ for
a) $\varepsilon_1 = 10.0\varepsilon_0$, $\varepsilon_2 = \varepsilon_0$, $\varphi_{1,\max} = 350°$ b) $\varepsilon_1 = 20.0\varepsilon_0$, $\varepsilon_2 = 10.0\varepsilon_0$, $\varphi_{1,\max} = 180°$ c) $\varepsilon_1 = -20.0\varepsilon_0$, $\varepsilon_2 = 10.0\varepsilon_0$, $\varphi_{1,\max} = 180°$. The inset depicts the geometry of the system (for more details see Fig. 5).

Another new phenomenon is revealed in Fig. 8, namely, near the ring junctions $\varphi = \pm\varphi_{1,\max}/2$ the electric flux is noticeably irregular with dips/spikes in the angular distribution. This new effect is caused by geometrical resonances of the structure, as explained next. Indeed, near $\varphi = \pm\varphi_{1,\max}/2$ there is a corner point common to both rings and also to the free-space region. On the one hand, the boundary conditions near this corner point impose that the azimuthal field $E_\varphi$ is continuous, because it is the tangential component of **E** with respect to the interface $r = R_2$. On the other hand, $E_\varphi$ is the normal component of **E** with respect to the rings junction $\varphi = \varphi_{1,\max}/2$, and thus it must be discontinuous at this interface. These two contradictory boundary conditions create an irregular behavior of the fields near the junction among three different materials, which is the main cause of the revealed dips/spikes in the angular characteristic of the flux.

Obviously, these effects are undesirable and are difficult to take into account an equivalent circuit model. Another, secondary reason for the observed irregularity of the flux near the junctions is the possible excitation of surface plasmon polaritons (SPP) near these interfaces, which may occur when the permittivities of the two rings have opposite signs.

In order to get around these mentioned constraints, in the following sections we will introduce the concept of optical nanoinsulators – which may help minimizing the flux leakage – and the concept of optical nanoconnectors – which may help reducing the effect of the geometrical resonances discussed above.

## III. Optical nanoinsulators

From the results of the previous section, it is apparent that, due to the flux leakage and the coupling between the nanoparticles and the surrounding background material, the performance of a straightforward realization of the envisioned optical nanocircuits may be distinct from that of their low frequency counterparts. The problem is that the displacement current $-i\omega\mathbf{D}$ induced in the nanowire does not need to be physically confined inside the material, distinctly from what happens at low frequencies in relatively good conductors where the drift path of the free-conduction charges is inherently bounded by the shape of the conductor

To circumvent these problems, we propose here to properly "shield" the optical nanoelements with a nanoinsulator for the displacement current in optical domain. In order to heuristically understand which materials may have the proper characteristics to behave as optical nanoinsulators, next we revisit the previously referred analogy/duality between our optical circuits and classical magnetic systems [22]. It is well-known, that in magnetic systems the magnetic flux induced in a magnetic core tends to be completely confined inside the circuit and that the leakage flux is residual. The justification of this phenomenon is very plain: the permeability of the magnetic core, $\mu$, is several orders of magnitude greater than that of the free-space region (a typical value is of about $\mu > 2000\mu_0$), and this huge permeability contrast forces the magnetic induction lines to be confined within magnetic core. How can we take advantage of this information to eliminate the flux leakage? One possibility, still exploring the analogy between our optical circuits and the classical magnetic circuits, is to impose the permittivity of the nanoelements to be much larger than that of the background material in absolute value, $|\varepsilon| \gg \varepsilon_0$. Under these circumstances, the electric flux leakage is expected to be small, as supported by the example with $\varepsilon = 100.0\varepsilon_0$ in Fig. 4. However, we may look for an alternative solution for which the nanocircuit elements can have moderate permittivity values.

To this end, we analyze more carefully the condition that ensures that the flux leakage is small: $|\varepsilon| \gg \varepsilon_0$. Evidently, if the nanoelement is covered with a material with permittivity $\varepsilon_{shield}$, instead of standing in free-space, the condition becomes, $|\varepsilon| \gg \varepsilon_{shield}$. The previous formula suggests that materials with permittivity near-zero, i.e., $\varepsilon_{shield} \approx 0$, may be suitable to work as optical nanoinsulators for the displacement current, analogous to what happens in classic circuits with the poor conductivity of the background materials. As referred in section I, these ENZ materials may be available in nature at terahertz, IR and optical frequencies when metals, semiconductors, and plasmonic materials are near their plasma frequency [9]-[13].

Before numerically testing this conjecture, let us show that the same conclusion regarding the properties of the nanoinsulator material may be obtained directly from the electromagnetic field theory. In fact, the displacement current through the side walls of a nanowire covered with a shield with permittivity $\varepsilon_{shield}$ is given by $\mathbf{J}_d = -i\omega\mathbf{D}$. Since the normal component of $\mathbf{D} = \varepsilon\mathbf{E}$ is continuous at a dielectric interface, it is clear that if $\varepsilon_{shield} \approx 0$ and if the electric field inside the ENZ-material remains finite in the $\varepsilon_{shield} = 0$ limit, then no displacement current can penetrate inside it. Therefore an ideal ENZ-material may behave as a perfect shield for the displacement current.

It is important to underline and stress that the proposed optical nanoinsulators are shields for the displacement current, but not shields for the electromagnetic field. That is, even though these nanoinsulators block the leakage of the displacement current, and thus forcing it to flow inside the nanocircuit element, a shielded nanoelement is not an isolated electromagnetic entity. Indeed, it can very well radiate and, eventually, couple some energy from the exterior. In some senses, as already outlined, these shielded nanoelements behave as conventional elements at low frequencies. In fact, also in regular conductors the current is completely confined inside the material volume, but indeed conducting wires may radiate and couple electromagnetic energy with the background.

### A. *Uniform nanoring shielded by a nanoinsulator*

In order to demonstrate the suggested possibilities for isolating nanocircuit elements, we go back to the same 2D geometry analyzed in the previous section. A uniform ring with permittivity $\varepsilon$ is fed by two balanced magnetic line sources. However, in order to block the flux leakage, the nanocircuit is now covered with two ENZ-nanoinsulators, as shown in panel *b* of Fig. 2. The thickness of the ENZ insulators is $\delta_{R,ENZ} = 0.05q\lambda_0$ and the line sources are positioned along the $x$-axis at $R_s^+ = 1.15q\lambda_0$ and $R_s^- = 0.65q\lambda_0$. As in section II, the dimensions of the ring are $R_1 = 0.8q\lambda_0$ and $R_2 = 1.0q\lambda_0$. The computed normalized flux inside the ring is shown in Fig. 9 as a function of $\varphi$, for $\varepsilon = 10\varepsilon_0$ (nanocapacitor) and $\varepsilon = -10\varepsilon_0$ (nanoinductor). The permittivity of the nanoinsulators at the design frequency was taken equal to $\varepsilon_{shield} = 0.001\varepsilon_0$ (solid lines). As seen in Fig. 9, the ENZ-shields effectively block the flux leakage, guiding the displacement current along the circuit path and forcing $\phi_e$ to be nearly constant inside the ring. When the permittivity of the shield is increased ten times, $\varepsilon_{shield} = 0.01\varepsilon_0$, the blockage of the displacement current is not as effective (dashed lines in Fig. 9). However, it is always possible to improve the insulating properties of the shield by increasing its thickness (this will be shown later in other configurations). Indeed, even for $\varepsilon_{shield} = 0.01\varepsilon_0$ the results are quite remarkable because the shields are extremely thin and the line sources are very close to the circuit path. An important consequence of these results is that the insulated nanowire may be accurately described by the circuit theory, more specifically by (2) and (3). For example, for the case $\varepsilon = 10\varepsilon_0$ and $\varepsilon_{shield} = 0.001\varepsilon_0$, the average flux calculated numerically is $\langle\phi_e\rangle/\varepsilon_0 = 0.357$ [V], which yields a (2D) electrical reluctance equal to $\Re_e = V/\phi_{e,av} = 2.80/\varepsilon_0$, whereas the result predicted by (2) is $\Re_e = 2.82/\varepsilon_0$ [F/m]$^{-1}$. In Appendix A we formally show how these conclusions hold in an exact way when the permittivity of the shield material tends to zero, in principle independent of its thickness.

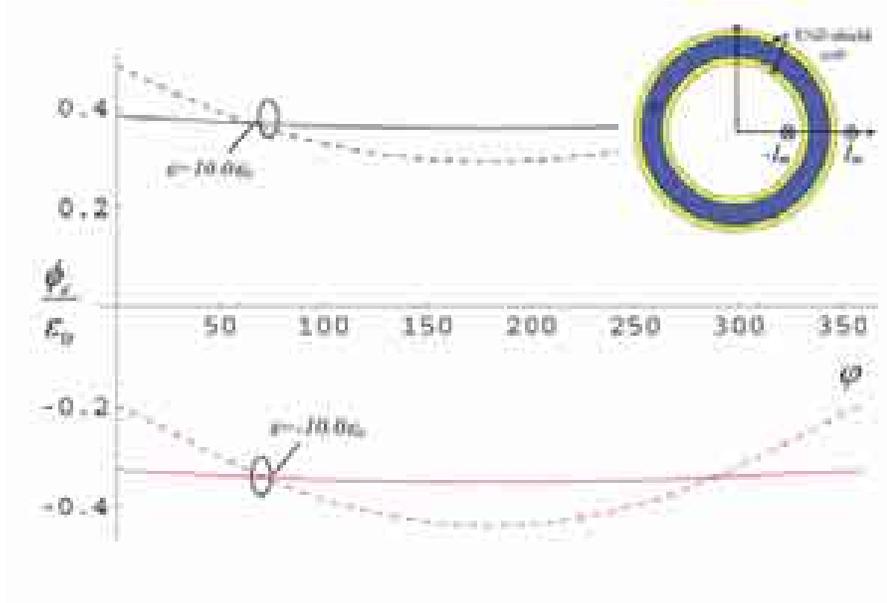

**Fig. 9.** (Color online) Normalized electric flux (p.u.l.) inside the subwavelength insulated ring as a function of $\varphi$ for $\varepsilon = 10.0\varepsilon_0$ and $\varepsilon = -10.0\varepsilon_0$. Solid lines: $\varepsilon_{shield} = 0.001\varepsilon_0$. Dashed lines: $\varepsilon_{shield} = 0.01\varepsilon_0$. The inset shows the geometry of the structure.

## *B. Parallel interconnection shielded by nanoinsulators*

The proposed optical nanoinsulators may not only minimize the interaction between the nanocircuit and the contiguous background region, but also reduce the undesired coupling between adjacent nanocircuit elements. To illustrate this effect we revisit the parallel circuit configuration, depicted in panel *a* of Fig. 5. However, now we assume that the rings are covered with two ENZ nanoinsulators, as shown in the inset of Fig. 10. The permittivity and thickness of the ENZ shields are those of the previous example, as well as the feed configuration. In the first example, we consider that the permittivity of the inner ring is $\varepsilon_{in} = 10\varepsilon_0$, the permittivity of the outer ring is $\varepsilon_{out} = 20\varepsilon_0$, and the radii of the rings are $R_1 = 0.8q\lambda_0$, $R_{int} = 0.9q\lambda_0$, and $R_2 = 1.0q\lambda_0$. The induced electric fluxes inside the two rings are depicted in Fig. 10 (curves labeled with symbol *a*). It is remarkable, that $\phi_{in}$ and $\phi_{out}$ become nearly constant and invariant with $\varphi$, in contrast to

-26-

what happens when the shields are removed (Fig. 6). The average values for the electric fluxes are $\langle \phi_{out} \rangle / \varepsilon_0 = 0.36$ and $\langle \phi_{in} \rangle / \varepsilon_0 = 0.20$ [V], and the electromotive force calculated numerically is $V = 1.09$ [V]. These values yield the reluctances $\Re_{out} = 2.98/\varepsilon_0$ [F/m]$^{-1}$ and $\Re_{in} = 5.32/\varepsilon_0$ [F/m]$^{-1}$, in excellent agreement with the circuit model (5)-(6) developed in section II. In the second example, we considered that $\varepsilon_{in} = 10\varepsilon_0$ and $\varepsilon_{out} = -10\varepsilon_0$. The numerically calculated $\phi_{in}$ and $\phi_{out}$ are depicted in Fig. 10 (solid curves labeled with symbol *b*). Despite the use of the two nanoinsulators, some variation of the fluxes with the azimuthal angle is still noticeable, which is mainly due to an exchange of current between the parallel elements. However, as compared to the results of the unshielded case (Fig. 7) the improvement is noticeable. In fact, in section II it was demonstrated that in the unshielded case the reported strong flux oscillations are related to the excitation of SPPs at the interface between the inner and outer rings. As seen in Fig. 10 the use of nanoinsulators prevents the excitation of SPPs, and greatly improves the confinement of the displacement current inside the circuit path. Even better insulation may be obtained by either considering shields with permittivity closer to zero or by increasing the thickness of the ENZ-shields. This is also illustrated in Fig. 10, where we plot the induced fluxes when the thickness of the ENZ-shields is increased four times (dashed curves labeled with symbol *b*; for this example the position of the line sources is $R_s^+ = 1.3q\lambda_0$ and $R_s^- = 0.5q\lambda_0$). Consistently with our intuition, it is seen that the flux becomes more uniform and nearly constant inside the two rings.

It is remarkable that the ENZ-nanoinsulators were able to prevent the excitation of SPPs, even though we did not place an ENZ-nanoinsulator in between the two rings (which however would further enhance the performance of this parallel circuit, completely isolating the two elements). The formal justification of these effects is given in Appendix B.

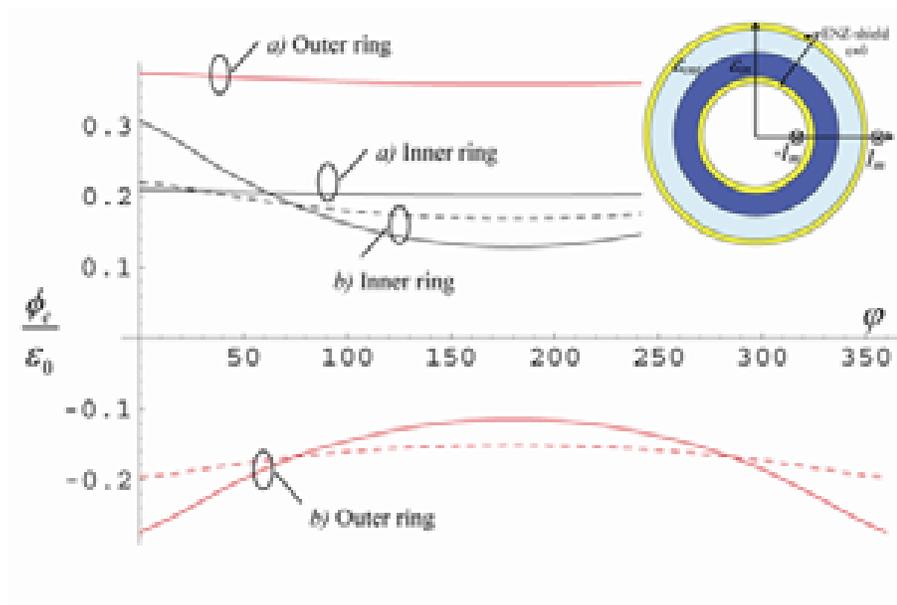

**Fig. 10.** (Color online) Normalized electric flux (p.u.l.) inside the subwavelength insulated rings as a function of $\varphi$ for a) $\varepsilon_{in} = 10.0\varepsilon_0$ and $\varepsilon_{out} = 20.0\varepsilon_0$ b) $\varepsilon_{in} = 10.0\varepsilon_0$ and $\varepsilon_{out} = -10.0\varepsilon_0$. The permittivity of the ENZ-shield is $\varepsilon_{shield} = 0.001\varepsilon_0$. The inset shows the geometry of the structure. The dashed lines correspond to case b) with ENZ-shields four times thicker.

### C. Series interconnection shielded by nanoinsulators

It is also pertinent and instructive to assess the effect of the envisioned optical nanoinsulators in the series circuit configuration depicted in panel *b* of Fig. 5. It was seen in section II that a straightforward realization of the series arrangement as the simple cascade of two nanocircuit elements may behave differently from what is expected from a conventional circuit theory, and in particular it was seen that the induced displacement current may not be confined within the circuit path, and that instead it may leak out to the adjoining region (Fig. 8).

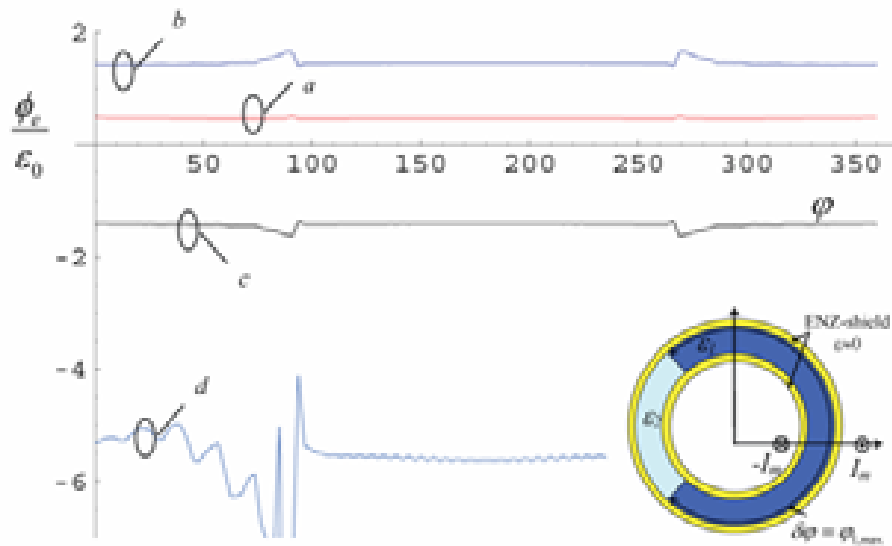

**Fig. 11.** (Color online) Normalized electric flux (p.u.l.) inside the shielded subwavelength ring as a function of $\varphi$ for *a)* $\varepsilon_1 = 10.0\varepsilon_0$, $\varepsilon_2 = 20\varepsilon_0$ *b)* $\varepsilon_1 = 10.0\varepsilon_0$, $\varepsilon_2 = -20.0\varepsilon_0$ *c)* $\varepsilon_1 = -10.0\varepsilon_0$, $\varepsilon_2 = 20.0\varepsilon_0$ *d)* $\varepsilon_1 = -9.0\varepsilon_0$, $\varepsilon_2 = 10.0\varepsilon_0$. The inset depicts the geometry of the system. In all the examples $\varphi_{1,\max} = 180^\circ$.

In order to analyze if the proposed nanoinsulators may help improving this situation, we have enclosed the subwavelength ring within two ENZ-shields, as depicted in the inset of Fig. 11. The dimensions and material properties of the shields, as well as the line source configuration, are the same as in section III.A. The rings are defined by $R_1 < r < R_2$, with $R_1 = 0.8q\lambda_0$ and $R_2 = 1.0q\lambda_0$. In Fig. 11 the electric flux inside the ring is shown as a function of $\varphi$, for different values of the material permittivities $\varepsilon_1$ and $\varepsilon_2$. In all the examples we have assumed that the ring sector with permittivity $\varepsilon_1$ is defined by $|\varphi| \leq \varphi_{1,\max}/2$, with $\varphi_{1,\max} = 180[\deg]$. In contrast with the results of Fig. 8, it is seen that the induced flux becomes nearly uniform and independent of $\varphi$, particularly for the set of parameters labeled with the symbols *a*), *b*) and *c*). For example, for $\varepsilon_1 = 10.0\varepsilon_0$ and $\varepsilon_2 = 20\varepsilon_0$ (curve *a* in Fig. 11, which corresponds to two nanocapacitors in series), we calculated numerically that the average flux p.u.l. is $\langle \phi_e \rangle / \varepsilon_0 = +0.482$ [V], which yields the equivalent reluctance $\Re_{eq} = 2.07/\varepsilon_0$ [F/m]$^{-1}$. This value agrees well with the theoretical formula (8), which gives $\Re_{eq} = \Re_1 + \Re_2 = (1.41/\varepsilon_0 + 0.71/\varepsilon_0)$[F/m]$^{-1} = 2.12/\varepsilon_0$ [F/m]$^{-1}$. Furthermore, we numerically calculated the electromotive forces $V_1$ and $V_2$ induced along the two nanocapacitors. We found that $V_1 = 0.66$ [V] and $V_2 = 0.35$ [V], while the theoretical values predicted by circuit theory (7) are $V_1 = 0.67$ [V] and $V_2 = 0.33$ [V]. These results clearly show how, by insulating the nanocircuit with ENZ-shields, it may be possible to describe the electrodynamics of the structure using classic circuit theory, providing the possibilities for design of more complex nanocircuits at optical wavelengths. Similar results and conclusions are obtained for the configurations *b*) and *c*). In the last example, we have simulated the same series arrangement for rings with $\varepsilon_1 = -9.0\varepsilon_0$, $\varepsilon_2 = 10.0\varepsilon_0$ (curve *d* in Fig. 11). Unlike the other examples, here the flux $\phi_e$ has noticeable fluctuations near the junctions of the two materials ($\varphi = \pm 90$ [deg]), even though away from the junctions the flux is, to some extent, uniform, apart from some visible ripple. The reason for this observed behavior is that the set of parameters $\varepsilon_1 = -9.0\varepsilon_0$, $\varepsilon_2 = 10.0\varepsilon_0$ corresponds to an LC series configuration close to its resonance, since effectively a nanoinductor ($\varepsilon_1 = -9.0\varepsilon_0$) has been placed in series with a nanocapacitor ($\varepsilon_2 = 10.0\varepsilon_0$). The total impedance associated with this arrangement is very small (or in other words, the equivalent impedance is near zero, since a resonant series configuration looks like a short circuit), and thus the amplitude of induced flux is relatively large, as is apparent from Fig. 11. Due to this resonant behavior, the quasi-static circuit theory may be limited and inadequate in describing all the peculiarities of the phenomenon (also the numerical accuracy of the MoM simulations may be somehow affected by this resonance). Other effects that may also play a role here are the geometrical and polaritonic resonances identified in section II that occur near the junction of the rings. The fact that the equivalent impedance of the nanocircuit is almost zero may exaggerate these phenomena, which cause the irregular behavior of the induced displacement current near the junctions. In the following section we suggest a strategy to minimize these unwanted localized resonances by using properly designed optical "nanoconnectors" placed at the location of the resonant interfaces.

For sake of completeness, we derive in Appendix C the quasi-static solution of this series problem in the limit of $\varepsilon_{shield} = 0$.

## *D. Modeling a realistic feed for the optical circuit*

So far in our computational models we have used two balanced magnetic line sources as the feeding mechanism of the proposed optical circuits. From a computational and conceptual point of view this choice is very convenient since it can be easily implemented in the MoM numerical code, and also because it is a clean and simple way of imposing a known electromotive force along the nanocircuits under study. Unfortunately, as pointed out before, from a practical perspective such feeding mechanism is unrealistic, since there are no magnetic charges in nature.

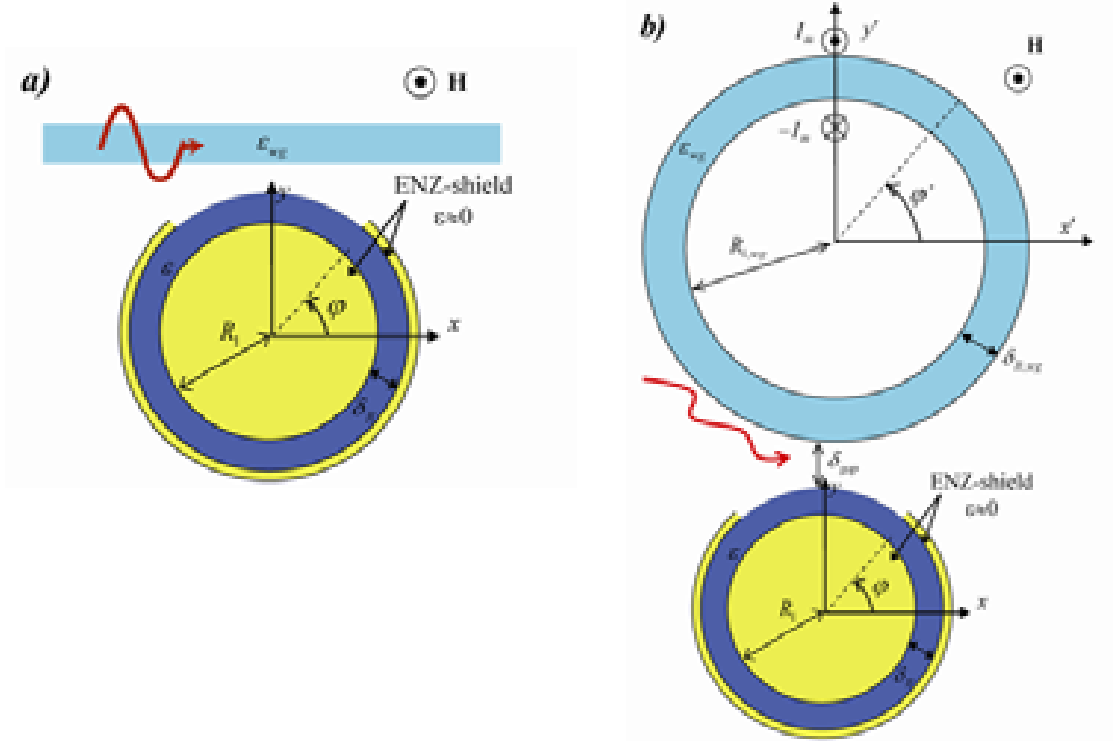

**Fig. 12.** (Color online) *a*) Configuration proposed to couple the electromagnetic energy guided by a slab waveguide with permittivity $\varepsilon_{wg}$ to a shielded nanocircuit. *b*) Configuration implemented in our MoM code in order to ensure that the computational domain is finite.

Our objective here is to propose a simple and more realistic excitation mechanism to feed the nanocircuit. More specifically, we suggest to couple the fields guided by a slab waveguide to the nanocircuit, as illustrated in panel *a* of Fig. 12. Our intuition and expectation is that the incoming wave will induce an electromotive force in the vicinity of the ring, feeding the nanocircuit in this way. For the sake of simplicity, we assume that the geometry is two-dimensional and uniform along the *z*-direction. As shown in Fig. 12*a*, the shielded subwavelength ring is illuminated by a guided mode that propagates tightly attached to an infinite slab waveguide with permittivity $\varepsilon_{wg}$. Note that the exterior ENZ-shield does not surround completely the whole ring, leaving an uncovered sector near the slab to improve the electromagnetic coupling. The core of the subwavelength ring is completely filled with an ENZ-material, to prevent the flux leakage to the interior region. In order for the incoming wave to be tightly bounded to the waveguide and for the waveguide cross-section to be subwavelength (to ease the numerical simulations), we assume that the slab waveguide is made of a plasmonic material with permittivity $\varepsilon_{wg} = -2.0\varepsilon_0$ at the design frequency. As is well-known, such waveguide may indeed support guided sub-wavelength plasmonic modes that are intrinsically related to the excitation of surface plasmon polaritons at the interfaces between the waveguide and the background material.

The full wave simulation of the structure described in Fig. 12*a* using the MoM is a challenging task, since this numerical method cannot easily characterize unbounded structures (namely, the infinite slab waveguide). To circumvent this problem, we have simulated numerically the structure shown in panel *b* of Fig. 12, which we expect may mimic, to some extents, some of the features of the configuration shown in panel *a*. The idea is to replace the infinite slab waveguide by a large ring-shaped waveguide with permittivity $\varepsilon_{wg}$. Since the radius of curvature of this ring is much larger (in our simulations 5 times) than the radius of curvature of the optical circuit, the curved waveguide will look *locally* plane and interact with the nanocircuit nearly in the same way as a planar slab waveguide. As depicted in Fig. 12*b*, the curved waveguide is fed by the same balanced line source configuration used in previous examples. This will excite the surface wave mode that illuminates the nanocircuit. Notice that the balanced line source is only used here to excite the surface wave mode in the curved waveguide, but does not interact directly with the nanocircuit.

In our simulations we have assumed that the dimensions of the curved waveguide are $R_{1,wg} = 4.0q\lambda_0$ and $\delta_{R,wg} = 1.0q\lambda_0$, and that the line sources are positioned at $R_s^- = 3.25q\lambda_0$, $R_s^+ = 5.75q\lambda_0$ along the y'-axis, with $q = 0.002$ (see Fig. 12b). On the other hand, the nanocircuit consists of a ring with permittivity $\varepsilon$, and it is defined by $R_1 < r < R_2$, with $R_1 = 0.9q\lambda_0$ and $R_2 = 1.0q\lambda_0$. The core of the ring, $r < R_1$, is filled with an ENZ-material with $\varepsilon_{shield} = 0.01\varepsilon_0$, and the ring is partially enclosed by a shield with the same permittivity and thickness $\delta_{R,ENZ} = 0.1q\lambda_0$. As seen in Fig. 12b, the angular sector $90 - 45/2 < \varphi < 90 + 45/2$ [deg] is not insulated, to allow good coupling with the incoming wave. The gap between the curved waveguide and the nanocircuit is $\delta_{gap} = 0.2q\lambda_0$.

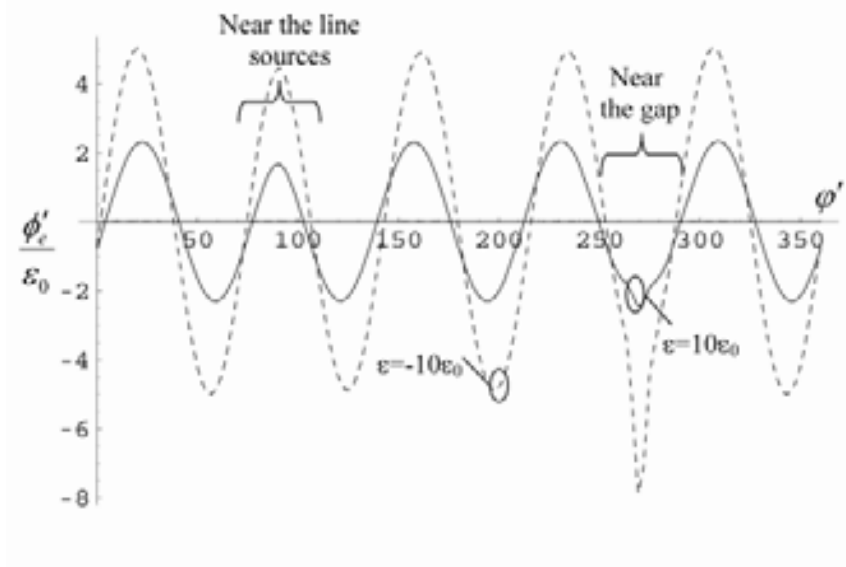

**Fig. 13.** Normalized electric flux (p.u.l.) inside the curved waveguide as a function of $\varphi'$ for $\varepsilon = 10.0\varepsilon_0$ (solid line) and $\varepsilon = -10.0\varepsilon_0$ (dashed line).

In Fig. 13 we plot the induced flux (p.u.l) inside the curved waveguide as a function of $\varphi'$ ($\varphi'$ is measured relatively to the coordinate system centered at the center of the curved waveguide, as shown in Fig. 12*b*) for different values of the permittivity of the nanocircuit. It is seen that the flux inside the curved waveguide is highly oscillatory consistently with our expectation that a surface wave is excited at the interfaces between air and the plasmonic material. As indicated in the figure, the region near $\varphi' = 90$ [deg] corresponds to the vicinity of the line sources, while the region near $\varphi' = 270$ [deg] corresponds to the vicinity of the nanocircuit.

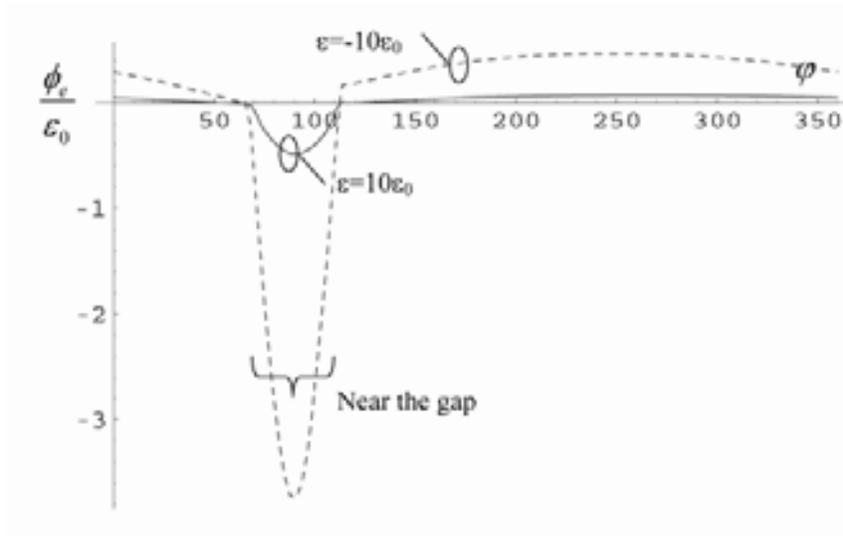

**Fig. 14.** Normalized electric flux (p.u.l.) inside the nanocircuit as a function of $\varphi$ for $\varepsilon = 10.0\varepsilon_0$ (solid line) and $\varepsilon = -10.0\varepsilon_0$ (dashed line).

In Fig. 14 the corresponding flux (p.u.l) variation along the nanocircuit is shown. Here the gap region corresponds to the vicinity of $\varphi = 90$ [deg]. Consistently with our expectations, it is seen that apart from the non-insulated region the induced flux is nearly uniform inside the nanocircuit. From a circuit point of view the gap region $90 - 45/2 < \varphi < 90 + 45/2$ [deg] may be interpreted as the "generator" or "battery" of the system. To better explain this concept, let us consider the case in which $\varepsilon = -10\varepsilon_0$ (equivalent circuit is a nanoinductor). We computed numerically for this case the (counterclockwise) electromotive force across the unshielded region, which turns out to be $V_{gap} = 1.75$ [V]. Consistently also with the quasi-static approximation, the voltage drop along the insulated region of the ring $-270 + 45/2 < \varphi < 90 - 45/2$ [deg] is $V_{circuit} \approx -V_{gap} = -1.75$ [V]. On the other hand, the reluctance of the insulated portion of the nanocircuit is, $\Re_e = \dfrac{\frac{7\pi}{4} R_{med}}{\varepsilon \delta_R} = \dfrac{\frac{7\pi}{4} 0.95}{-10\varepsilon_0 \times 0.1} = -\dfrac{5.22}{\varepsilon_0}$ [F/m]$^{-1}$ (which corresponds to a nanoinductance). Hence, using (3) one expects that the flux (p.u.l) induced inside the insulated section of the circuit is given by $\phi_e = \dfrac{V_{circuit}}{\Re_e}$, which yields $\dfrac{\phi_e}{\varepsilon_0} = +0.33$ [V]. This value is completely consistent with the results depicted in Fig. 14 (dashed line), where it is seen that $\dfrac{\phi_e}{\varepsilon_0}$ is relatively close to this theoretical value in the insulated section of the ring. In fact, numerical integration of the full wave simulation results also shows that the average value of the flux is $\left\langle \dfrac{\phi_e}{\varepsilon_0} \right\rangle = 0.33$ [V] over the shielded region $-270 + 45/2 < \varphi < 90 - 45/2$ [deg], which fully supports our circuit model. This simple example clearly shows how a realistic "voltage generator" at optical frequencies is within the realm of possibility, and how this voltage generator may be modeled and properly taken into account using the proposed circuit concepts at optical wavelengths even in more complex configurations. To conclude this section, and to give an idea of the field distribution in the problem studied here, we depict (Fig. 15) the amplitude of the magnetic field in the vicinity of the nanocircuit for the case $\varepsilon = 10.0\varepsilon_0$. It may be seen that the magnetic field has a maximum near the unshielded region showing the transfer of power between the curved waveguide and the nanocircuit.

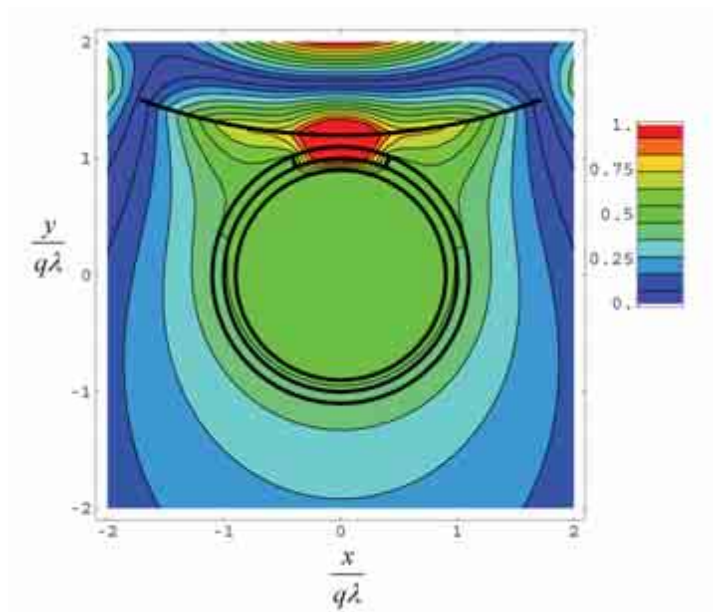

**Fig. 15.** (Color online) Contour plot for the normalized magnetic field in the vicinity of the insulated nanocircuit for the case $\varepsilon = 10.0\varepsilon_0$.

## *E. Simulations of three-dimensional arrangements of nanocircuit elements*

In this section (and in Section V), we confirm that the proposed nanocircuit concepts are not limited to two-dimensional structures and specific polarization of the field, but they may indeed be applied as well to the more realistic three-dimensional (3D) configurations of nanoparticles. To this end, we used the commercial finite-integration technique electromagnetic simulator CST Studio Suite[TM] [24] to characterize 3D arrangements of nanocircuit elements. In our simulations the nanoparticles are straight cylinders directed along the *z*-direction. We used a very simple excitation mechanism to impose a desired electromotive force across the nanowires. First of all, supposing that such nanoparticles are included within the region $0 < z < L$, we placed perfectly electric conducting (PEC) planes at $z=0$ and $z=L$. Then, using the functionalities of CST Studio Suite[TM] [24], we connected an ideal voltage source across the referred PEC planes. The voltage source is placed relatively far from the nanowires in order to avoid unwanted interferences. This simple configuration forces (in the quasi-static limit) the electromotive force to be nearly constant between the PEC plates, effectively imposing the prescribed voltage drop across the nanoelements under study. Of course, the described feeding mechanism is not realistic, but nonetheless it is appealing from the computational point of view for its simplicity, and, most importantly, it is sufficient to numerically characterize the effect of the nanoinsulators in relatively complex 3D-circuit setups. A more realistic form of excitation in 3D configuration is analyzed in Section V.

In the first example, we simulated an LC series arrangement of two nanocylinders with permittivities (at the frequency of interest) $\varepsilon_1 = 10\varepsilon_0$ (nanocapacitor) and $\varepsilon_2 = -15.0\varepsilon_0$ (nanoinductor). The nanowires are directed along $z$ and have circular cross-section with radius $R = 0.01\lambda_0$. The nanoparticle with permittivity $\varepsilon_1$ is defined from $0 < z < 0.4L$, and the nanowire with permittivity $\varepsilon_2$ is defined from $0.4L < z < L$, with $L = 5R$. The induced electric field vector distribution (snapshot in time) is shown in panel $a$ of Fig. 16 for a transverse cut of the nanocircuit. In Fig. 17 we plot the amplitude of the electric field component $E_z$ (dashed line) along the axis of the nanowires. The electric field is normalized to $E_0 = E_z(z=0)$. It is clearly seen that the displacement current is not confined inside the nanocircuit, and the leakage is well visible in Fig. 16. As a consequence it is seen in Fig. 17 (dashed line) that the electric field inside the nanowires is not uniform.

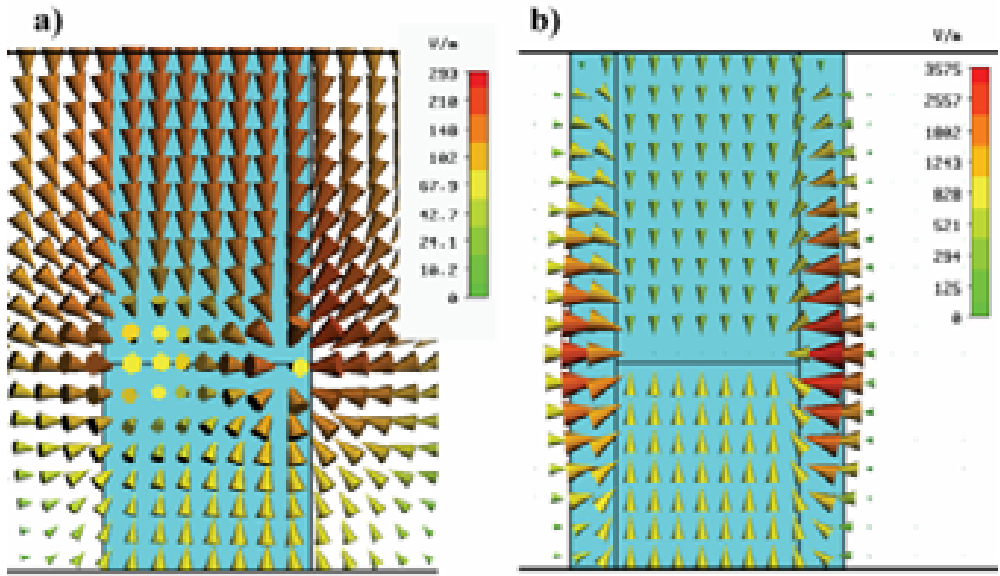

**Fig. 16.** (Color online) *a*) Snapshot in time of the electric field vector on the plane $\varphi = 0$ at the center of the nanocircuit. The nanowires are arranged in an LC series configuration. The lower region $0 < z < 0.4L$ has permittivity $\varepsilon_1 = 10\varepsilon_0$ (nanocapacitor) and the upper region $0.4 < z < L$ has permittivity $\varepsilon_2 = -15.0\varepsilon_0$ (at the frequency of interest) (nanoinductor). *b*) Same as panel *a*) but nanowires are insulated with an ENZ-material.

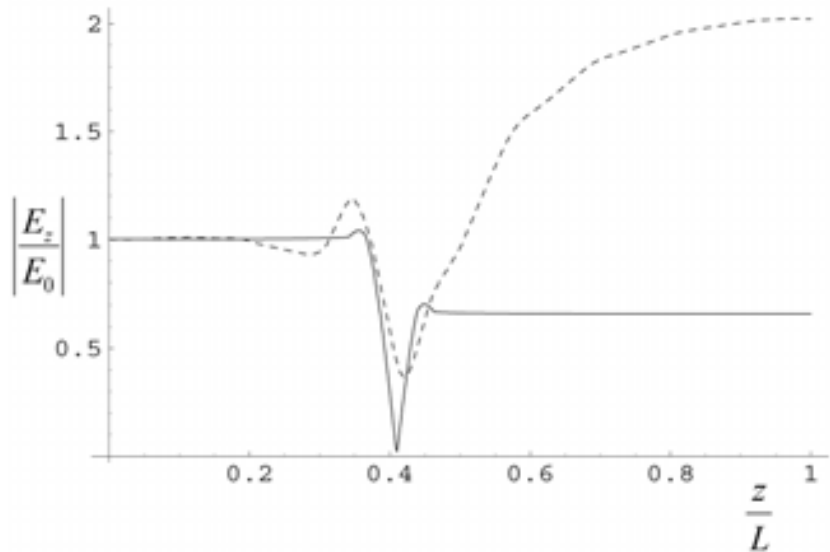

**Fig. 17.** Normalized electric field component $E_z$ along the axis of the nanowires for the configuration depicted in Fig. 16. Solid line: insulated nanoelement; Dashed line: nanoelement without shield.

Consistently with our expectations, the situation changes completely when the elements are insulated with an ENZ material. This case is reported in panel *b* of Fig. 16, where it is assumed that the radius of the ENZ-shield is $R_{shield} = 1.5R$, and that the ENZ material follows a Drude type model $\varepsilon = \varepsilon_0 \left(1 - \frac{\omega_p^2}{\omega(\omega + i\Gamma)}\right)$, where $\omega_p$ is the plasma frequency and $\Gamma$ is the collision frequency [rad/s]. The field distribution of Fig. 16 was calculated for $\omega = \omega_p$ and using $\Gamma = 0.01\omega_p$, so that the effect of mild realistic losses is considered (note that at $\omega = \omega_p$, we have $\text{Re}(\varepsilon) \approx 0$ and $\varepsilon/\varepsilon_0 \approx +i\,\Gamma/\omega_p$). Consistently with the results of the previous sections, it is seen that the ENZ shield effectively confines the electric displacement flux inside the nanocircuit elements. Also in Fig. 17 (solid line), it is seen that apart from the dip near the junction, the electric field is nearly uniform both inside the nanoinductor and the nanocapacitor, consistently with what expected from our circuit analogy. These results once again fully support our theoretical models, namely formula (7).

In the second example, the nanoelements are arranged in an LC parallel configuration. The nanowires are concentric this time, as seen in panel *a* of Fig. 18, and are defined from $0 < z < L$. The radii of the inner and outer nanowires are $R_{in} = 0.5R$ and $R_{ext} = R$, respectively, with $R = 0.01\lambda_0$ as in the previous example. The permittivity of the interior nanowire is $\varepsilon_{in} = 10\varepsilon_0$ (nanocapacitor) and that of the outer one is $\varepsilon_{out} = -15.0\varepsilon_0$ (nanoinductor). The electric field lines along a transverse cut of the nanocircuit is shown in panel *a* of Fig. 18. It is seen that differently from what happens in the series arrangement, the electric field is nearly uniform inside the nanowires, even though the wires were not insulated with an ENZ-material. In fact, due to the symmetries of our computational model it is not possible to excite SPPs at the interface between the nanowires, and consequently the circuit theory concepts apply here even without the use of insulating shields. In particular, (5)-(6) may be used to accurately characterize the LC parallel configuration.

In Section V, we will describe more realistic excitation of optical lumped nanocircuits by a plane wave.

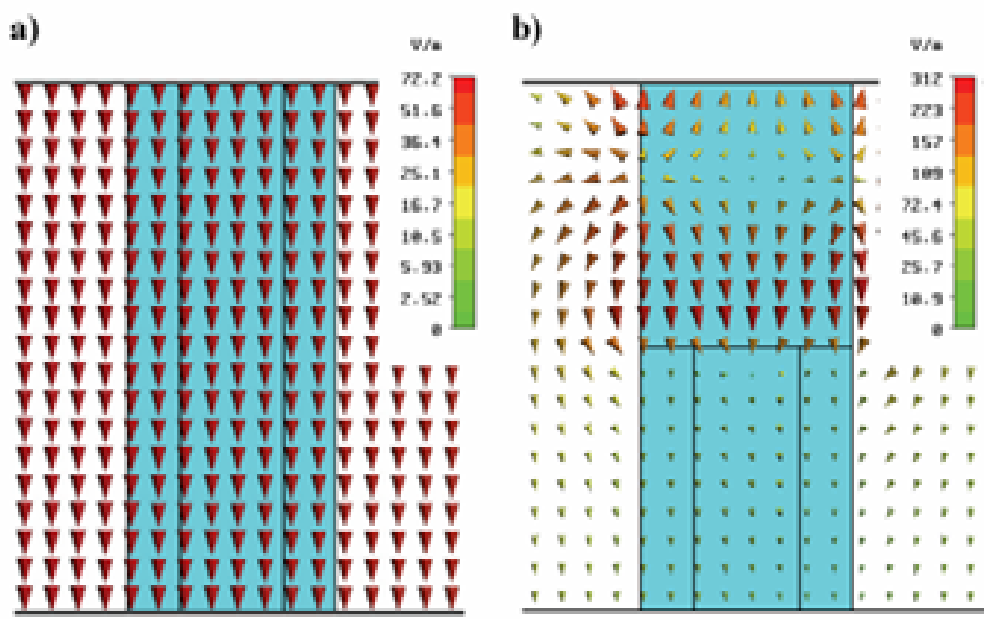

**Fig. 18.** (Color online) Snapshot in time of the electric field vector on the plane $\varphi = 0$ at the center of the nanocircuit. *a*) LC parallel configuration. *b*) Series of a nanoinductor (top section) with the parallel combination of a nanoinductor and a nanocapacitor (two concentric rods in the lower section).

## IV. Optical nanoconnectors

In the previous sections it was shown that the electric field near the junction of two nanoelements may become somehow irregular, due to geometrical and polaritonic resonances that may emerge at the interfaces between the materials. In particular, for the series combination of two nanoelements the induced displacement current may vary appreciably near the junction of the nanowires, as the spikes and dips of Fig. 8, Fig. 11, and Fig. 17 clearly demonstrate, even if the nanowires are properly insulated with an ENZ-shield. As noticed in section II, this effect is due to the singular nature of the electromagnetic fields in the vicinity of the intersection point of three dielectrics, as a consequence of conflicting boundary conditions. As presented analytically in [6], in fact, at such singular points the quasi-static potential distribution necessarily has a saddle point. Obviously, this irregular behavior is undesirable for practical purposes, since it may limit the applications of the proposed nanocircuit concepts. To further demonstrate the difficulties caused by this effect, we used CST Studio Suite$^{TM}$ [24] to simulate a (3D) nanocircuit configuration that consists of the series combination of a nanoinductor ($\varepsilon_{top} = -3\varepsilon_0$, top section in panel $b$ of Fig. 18, $0.5L < z < L$) with the parallel combination of a nanocapacitor and a nanoinductor (lower section in panel $b$ of Fig. 18, $0 < z < 0.5L$; the material parameters, radii of the two concentric nanowires and $L$ are the same as in the last example of section III.E). The feeding mechanism is also the same as in section III.E. The computed electric field lines are depicted in panel $b$ of Fig. 18. The polaritonic resonances near the interfaces are well visible, as well as the flux leakage. This is further supported by Fig. 19 and Fig. 20, where we plot (dashed lines) the normalized electric field component $E_z$ along the line segment $r = 0$ ($r$ is the radial distance with the respect to the axis of the nanowires) and $r = (R_{in} + R_{ext})/2$, respectively. Notice that for $0 < z < 0.5L$, the path $r = 0$ is inside the nanocapacitor with permittivity $\varepsilon_{in} = 10\varepsilon_0$, and the path $r = (R_{in} + R_{ext})/2$ is inside the nanoinductor with permittivity $\varepsilon_{out} = -15.0\varepsilon_0$. In order to avoid this undesirable current leakage, we have enclosed the nanocircuit in an ENZ shield with the same dimensions and material

-42-

properties as in the first example of section III.E. The corresponding electric field lines are shown in panel *a* of Fig. 21. The improvement as compared to the unshielded case is quite significant. This is also confirmed by Fig. 19 and Fig. 20, which show (solid black lines) that the electric field along the line segments $r = 0$ and $r = (R_{in} + R_{ext})/2$ is now more uniform than in the previous configuration. However, $E_z$ does still vary markedly near the junction ($z = 0.5L$) and also in the top nanowire $0.5L < z < L$. The origin of this phenomenon is related to the previously referred geometrical/polaritonic resonances near the junction between the three materials. As clearly seen in this example, these resonances may cause (from a circuit point of view) a poor physical "connection" between the nanoelements.

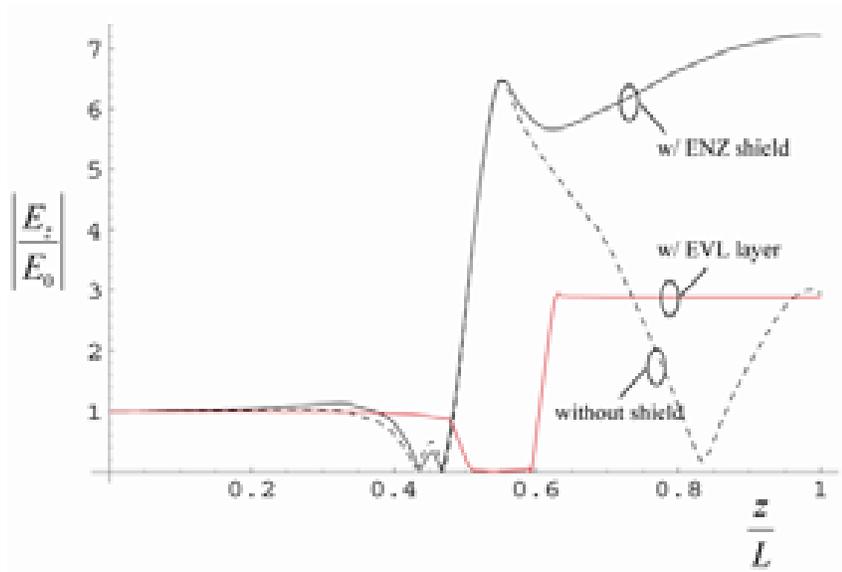

**Fig. 19.** (Color online) Normalized electric field component $E_z$ along the axis of the nanowires ($r = 0$) for the series interconnection of a nanoinductor with the parallel combination of a nanoinductor and a nanocapacitor. Solid black line: insulated nanocircuit; Dashed line: nanocircuit without shield; Solid red (lighter) line: insulated nanocircuit with an EVL connecting layer at the junction.

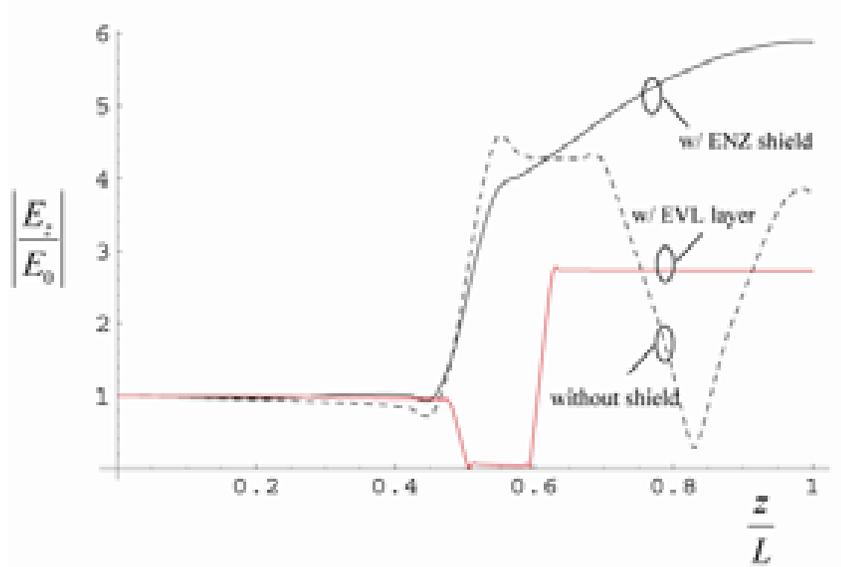

**Fig. 20.** (Color online) Same as Fig. 19 but the field is calculated along the line segment $r = (R_{in} + R_{ext})/2$.

How can we improve the connection between these nanowires (i.e., nanocircuit elements)? Which material can play the same role as good conductors at low frequencies and ensure a good circuit connection between the different lumped nanoelements? To answer these questions we note that at the RF and microwave frequencies, good conductors can carry a large electric conduction current with a small applied voltage drop. Using (3) it is evident that the counterpart of these materials within the framework of optical nanocircuits are nanoparticles characterized by a near zero impedance (or equivalently near-zero reluctance $\Re_e \approx 0$). From (2), it is clear that these nanoelements may be materials with $\varepsilon$ very large (EVL) (plasmonic or nonplasmonic), or more generally materials with $\varepsilon$ relatively very large as compared to the other materials used to synthesize the nanocircuits. Thus, we propose to use these EVL materials as connecting layers of the envisioned nanocapacitors and nanoinductors. In fact, we expect that, provided the dielectric contrast between the EVL layer and the other materials at the junction is sufficiently high (let us say 10 times), the effect of the previously mentioned geometrical and polaritonic resonances at the junctions will be strongly reduced. Note that some metals and some polar dielectrics behave naturally as EVL materials at IR and optical frequencies, even though they do not necessarily behave as good conductors for the conduction currents. In other words, they may act as good "optical conductors" for the displacement current in our circuit analogy (since the real parts of their permittivities can be relatively high), even though their conventional conductivity for the conduction current can be low (since the imaginary parts of their permittivities can be relatively low). Once again the role of conductivity for classic circuit concepts is played by the material permittivity in the present nanocircuit analogy at optical wavelengths.

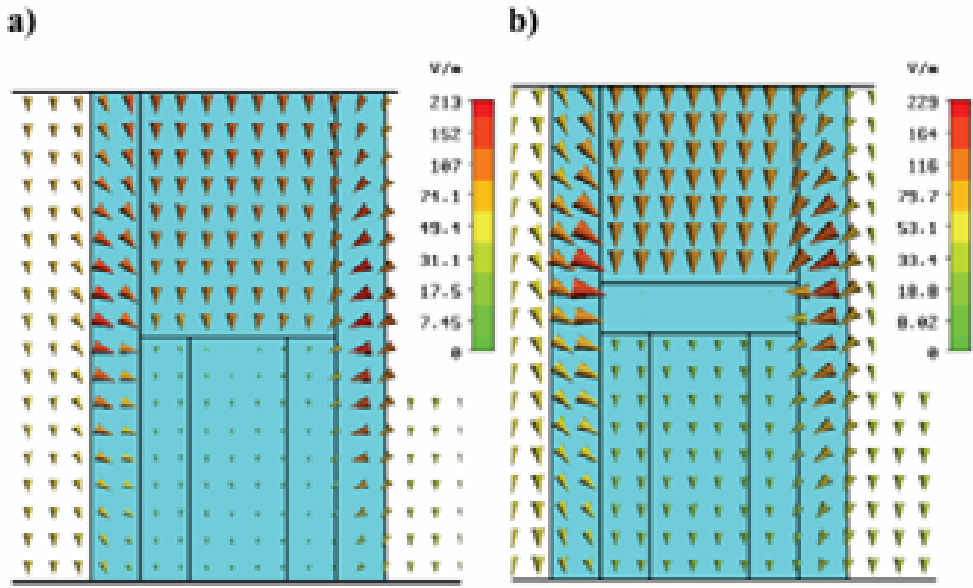

**Fig. 21.** (Color online) Snapshot in time of the electric field vector on the plane $\varphi = 0$ at the center of the nanocircuit. *a*) Series of a nanoinductor (top section) with the parallel combination of a nanoinductor and a nanocapacitor (two concentric rods in the lower section). The circuit is insulated with an ENZ material (exterior cylindrical layer). *b*) Same as *a*) but an EVL connecting layer is placed at the junction.

To test the proposed optical nanoconnector concept, we inserted an EVL layer at the junction between the nanoelements studied in the previous example. The assumed permittivity of the EVL layer is $\varepsilon_{EVL} = 200\varepsilon_0$, and is defined from $0.5L < z < 0.6L$ (the nanoinductor with permittivity $\varepsilon_{top} = -3\varepsilon_0$ is now defined from $0.6L < z < L$). The electric field lines are shown in panel *b* of Fig. 21. It is seen that the electric field inside the EVL layer is almost zero, and thus the optical voltage drop across the EVL material is practically zero, consistently with our heuristic interpretation that it may behave as a "nanoconnector". This property is supported by Fig. 19 and Fig. 20, which show the electric field inside the nanowires (solid red lines). Remarkably, the electric field becomes nearly constant inside the three nanowires, showing that the EVL layer effectively connects the different branches (i.e., different lumped nanoelements) of the optical nanocircuit. In particular, it has been demonstrated that the envisioned optical nanocircuits may be described even more consistently using the circuit theory, provided that the proposed lumped nanoelements are properly connected using an EVL components and properly insulated with ENZ materials.

## V. Complex 3-D Optical Nanocircuits

In this section using full-wave simulations we verify in more complex scenarios the theoretical and numerical results outlined in the previous sections, analyzing the electromagnetic behavior of 3-D optical nanocircuits in series and parallel configurations. To this end, we have simulated with CST Studio Suite 2006$^{TM}$ [24] several geometries involving 3-D sub-wavelength nanocircuits under plane wave excitation, which may model more thoroughly a realistic feed (e.g., an optical beam or a local NSOM excitation). The purpose of this study is also to analyze the behavior of such nanocircuit elements as a function of frequency, since the simulation allows us to fully take into account the material dispersion of ENG or ENZ materials, which is a necessary characteristic for such materials [2]. This may therefore represent a further step towards the full understanding of the frequency response of such nanocircuits, particularly for their potential use as optical lumped nanofiltering devices. Moreover, we fully take into account the possible presence of realistic absorption in these materials.

As a first set of simulations, we have studied the behavior of a 3-D nanocircuit composed of two nanoelements in series configuration, as depicted in Fig. 22a. The simulated structure consists of a nanocapacitor, made of a square cylinder (green, upper position in the figure) with side $l = \lambda_0 / 300$, with $\lambda_0$ being the background wavelength at the operating frequency $f_0$, and height $h = \lambda_0 / 100$, made of a dielectric material with $\varepsilon = 3\varepsilon_0$, connected to a nanoinductor of same size (light blue, lower position in the figure) made of an ENG material with permittivity following the Drude dispersion model $\varepsilon_{ENG}(\omega) = \varepsilon_0 \left(1 - \frac{(4\pi f_0)^2}{\omega(\omega + i\Gamma)}\right)$, where we have assumed for the damping radian frequency the value $\Gamma = 4\pi \cdot 10^{-2} f_0$, which is consistent with some values for optical plasmonic materials. The two nanoelements are connected with EVL nanoconnectors (darker blue) with permittivity $\varepsilon = 200\varepsilon_0$ and the whole nanocircuit is isolated with ENZ nanoinsulator shields (transparent) modeled with the Drude dispersion $\varepsilon_{ENZ}(\omega) = \varepsilon_0 \left(1 - \frac{(2\pi f_0)^2}{\omega(\omega + i\Gamma)}\right)$. The thickness of nanoconnectors and nanoinsulators has been fixed to $t = \lambda_0 / 600$ in this set of simulations. The whole nanocircuit, which indeed resembles a small portion of a lumped circuit, with the nanoconnectors acting as the "wires" connecting in series the nanocircuit lumped elements, is embedded in a background material with permittivity $\varepsilon_0$, and it is excited by a plane wave traveling along the positive *x* axis with electric field linearly polarized along *y* with amplitude 1 [V/m].

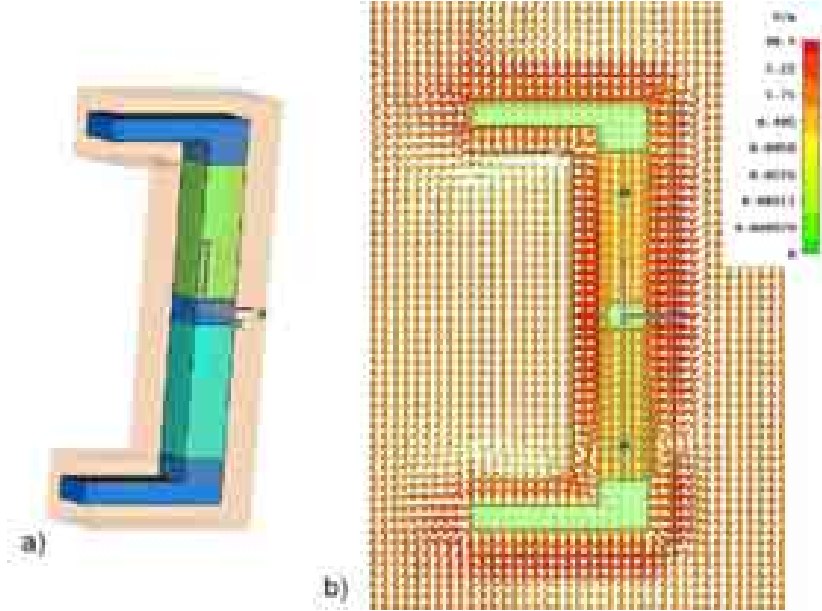

**Fig. 22.** (Color online) a) Geometry of a 3-D series nanocircuit formed by a nanoinductor (light blue, lower position) and a nanocapacitor (green, upper position) surrounded by ENZ nanoinsulators (transparent) and interconnected with EVL nanoconnectors (dark blue). b) Snapshot in time of the electric field vector induced on the plane $z = 0$ under plane wave incidence (with 1 [V / m] electric field amplitude).

Fig. 22b shows the electric field distribution (snapshot in time) on the symmetry plane cutting the nanocircuit (at $z = 0$) at the frequency $f_0$. Indeed, as expected, the optical "potential drop" in the nanoconnectors is very minor, due to the high permittivity of the EVL material, and indeed the electric field is oppositely directed in the two nanocircuit elements, ensuring that the equivalent optical displacement current flowing from one element to the other is the same, as required by the series interconnection between the nanoelements. We note that, although the nanocircuit is expected to be at the resonance (indeed the two permittivities at frequency $f_0$ are of opposite sign), the nanoparticles do not support a plasmonic resonance at their interface, due to the presence of the nanoconnector between them, as already discussed in the previous section. Indeed, in one of our simulations (not reported here) when the central nanoconnector was removed, the structure experienced strong unwanted plasmonic resonances at the interface between the two elements.

The electric field distribution in Fig. 22b shows another interesting feature: in the ENZ regions, due to the very low permittivity of these nanoinsulator regions, the field is orthogonal to the inner circuit and greatly enhanced, but still satisfying the boundary conditions for the displacement vector, which has to have negligibly small normal components at these different interfaces,. This again confirms our intuition regarding the analogy between such nanocircuits and the corresponding lower-frequency circuits. We also notice that the presence of the ENZ shield indeed stops the displacement current leakage out of the nanocircuit, but it does not necessarily nullify the presence of electromagnetic fields in such insulators (since the electric field is present in the ENZ regions, just as one would expect in a conventional plastic insulator in RF and microwave circuits.). As noticed in the previous sections, this is analogous to what happens in a conventional circuit element at lower frequencies, which may indeed radiate or couple energy with the surrounding, but indeed the background around it does not allow leakage of conduction current owing to very low (zero) conductivity of the background. Due to our different notion of current in the present nanocircuit analogy, the role of low conductivity materials in RF and microwave is taken by the low permittivity ENZ optical nanoinsulators here, which effectively eliminate the displacement current leakage from the nanocircuit and re-route this current along the path that is intended for. This may be further confirmed by other simulations we have performed (not reported here for sake of brevity), which simulated a different orientation of the electric field and propagation direction. Indeed, due to the presence of suitably designed nanoinsulators and nanoconnectors, the direction of the current flow inside the nanocircuit is weakly affected by the orientation of the exciting electric field and thus the current flow follows the same path reported in Fig. 22b even for skew incidence (although its magnitude may be different for different incident angles).

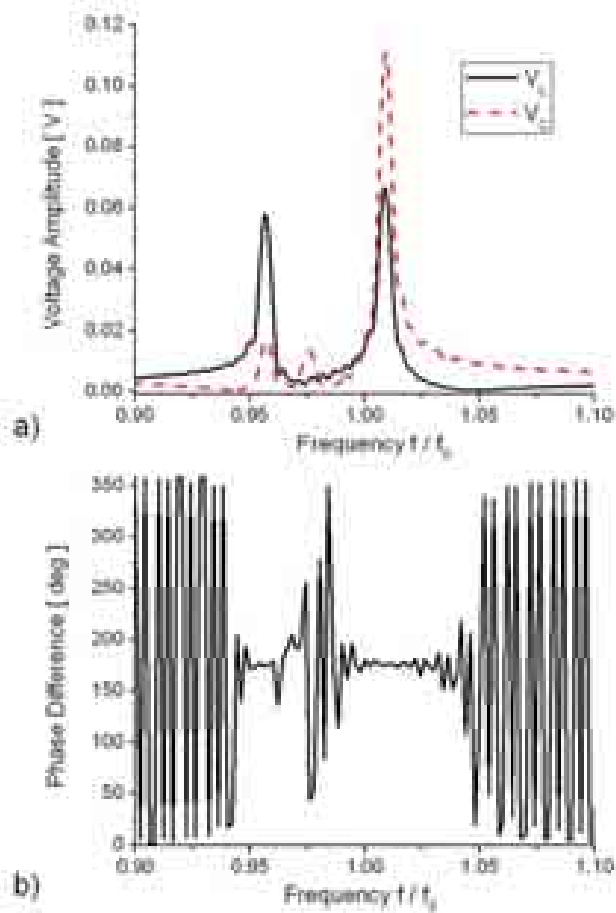

**Fig. 23.** (Color online) a) Optical voltage drop amplitude between the two ends of the nanoinductor (black solid line) and the two ends of nanocapacitor (red dashed line) of Fig. 22. b) Phase difference between the two voltage signals.

Fig. 23 shows the voltage amplitudes calculated across the two nanocircuit elements (along the long arrows in Fig. 22b) and the corresponding phase difference as a function of the frequency of operation normalized to $f_0$. It can be clearly seen how the voltage amplitudes experience a peak at $f_0$, due to the presence of a resonance in the nanocircuit, analogous to a classic series LC circuit. Moreover, the phase difference between the voltages is 180° at the operating frequency $f_0$, like in a series L-C circuit. This behavior is maintained over a relatively broad range of frequency, even though both the permittivity of the nanoinsulators and of the nanoinductor are frequency dispersive. These results are indeed quite promising for potential applications of these concepts and their feasibility as optical nanocircuits. (Outside this range of frequencies, the phase difference between the voltages across nanoelements differ from 180°, due to the fact that the permittivities of ENZ and ENG materials are different from what they have been designed for in the band around $f_0$. In particular, sufficiently away from its plasma frequency $f_p$, the insulator component acts as a DPS or ENG material, influencing the overall nano-circuit response. Nonetheless, the two components may still act as nanocircuit elements, albeit not necessarily as purely series LC.)

Fig. 24 shows the corresponding displacement current density across the two elements, calculated as the local relative permittivity (with respect to the background material) multiplied by the local electric field along the $y$ axis (long arrows in the Fig. 22b) at the frequency $f_0$. (For the sake of simplicity the multiplicative constant $-i\omega$ has been dropped from $-i\omega D$ term in this plot.) It is evident that the displacement current flow is almost constant across the two elements both in amplitude and in phase (in fact the local electric field is oppositely directed in the two nanoelements, consistently with Fig. 22b). The series connection between the elements is evident in this configuration, and it is indeed made possible by the presence of properly designed nanoinsulators and nanoconnectors.

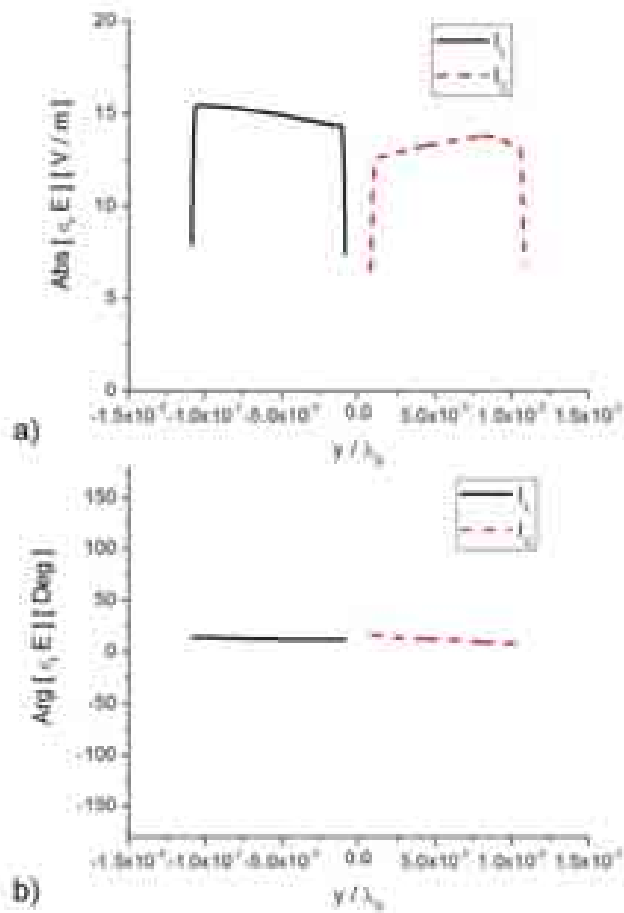

**Fig. 24.** (Color online) Variation of displacement current density amplitude (a) and phase (b) at frequency $f_0$, along the length of the nanoelements, calculated as the electric field amplitude at the center of each one of the two nanoelements multiplied by the corresponding relative permittivity, with respect to the background material, for the nanoinductor (black solid line) and nanocapacitor (red dashed line) of Fig. 22. (For simplicity, the multiplicative constant $-i\omega\varepsilon_0$ has not been included.)

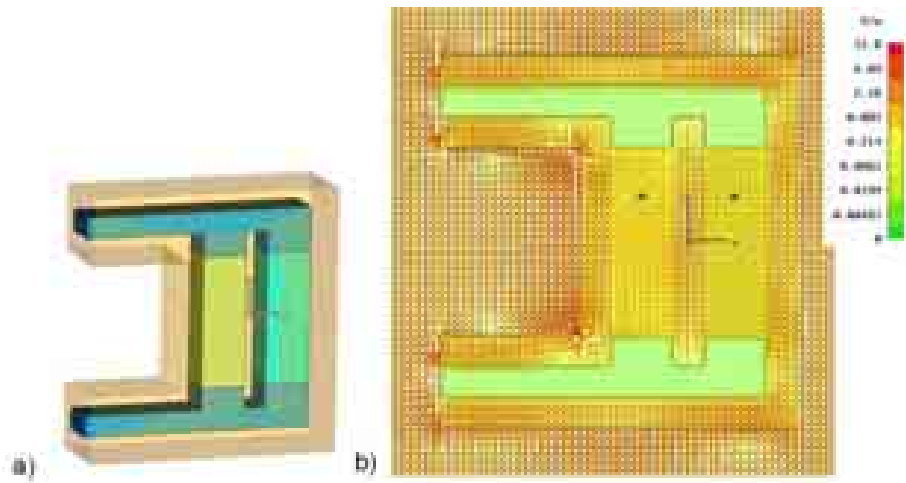

**Fig. 25.** (Color online) a) Geometry of a 3-D parallel nanocircuit formed by a nanoinductor (light blue, right position) and a nanocapacitor (green, left position) surrounded by ENZ nanoinsulators (transparent) and interconnected with EVL nanoconnectors (dark blue). b) Snapshot in time of the electric field vector induced on the plane $z = 0$ under plane wave incidence (with 1 [V / m] electric field amplitude).

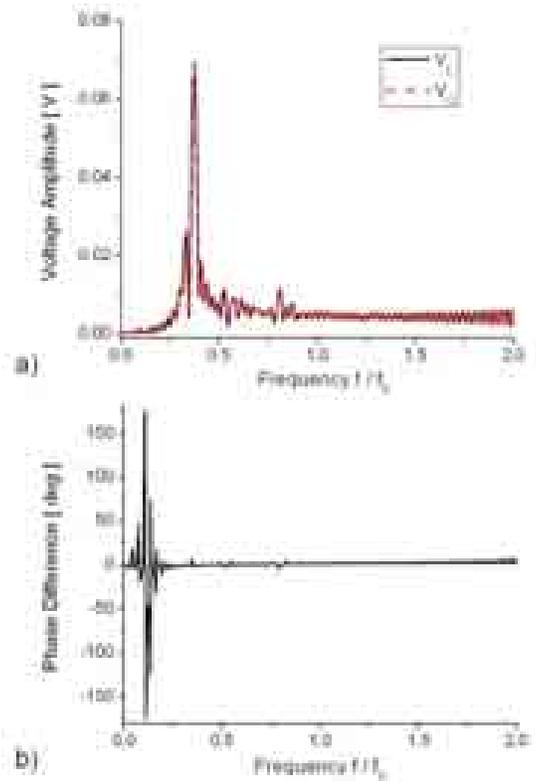

**Fig. 26.** (Color online) a) Optical voltage drop amplitude between the two ends of the nanoinductor (black solid line) and the two ends of nanocapacitor (red dashed line) of Fig. 25. b) Phase difference between the two voltage signals.

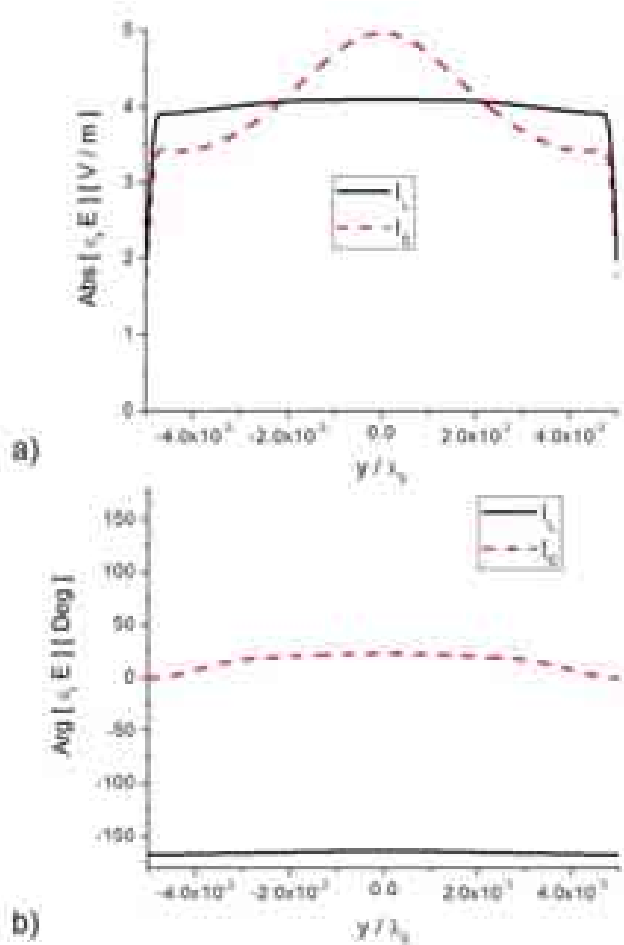

**Fig. 27.** (Color online) Variation of displacement current density amplitude (a) and phase (b) at frequency $f_0$ along the length of nanoelements, calculated as the electric field amplitude at the center of each one of the two nanoelements multiplied by the corresponding relative permittivity, with respect to the background material, for the nanoinductor (black solid line) and nanocapacitor (red dashed line) of Fig. 25. (For simplicity, the multiplicative constant $-i\omega\varepsilon_0$ has not been included.)

Fig. 25-27 report similar results for the parallel configuration for the same two nanocircuit elements. In this case the nanoconnectors have been properly modified in their geometry to excite the two nanoelements in parallel. The thickness $t$ of the nanoconnectors and nanoinsulators in this example has been kept the same as in Fig. 22. One can clearly see in this case that the electric field is parallel in the two elements, due to the fact that the optical voltage drops are in phase and the displacement current flows are opposite in phase (due to the opposite sign of permittivity), as expected in a parallel

L-C circuit. The voltage distribution versus frequency, reported in Fig. 26 shows how the two optical voltages are indeed very similar in amplitude and the phase difference between them is close to zero over a reasonably wide range of frequencies, even over a frequency range where the insulators are very far from behaving as displacement current shields. Fig. 27 reports the current densities across the two nanoelements at frequency $f_0$, also making evident the parallel interconnection between the nanoinductor and the nanocapacitor, with the current flow being quasi-uniform across the elements and the phase difference between them being around 180°. Two minor features of these plots might appear not to play in favor of our circuit analogy at the first look: the non perfect uniformity of the current density across the nanocapacitor and the absence of a peak in the voltage distribution at the resonance of the system, which is supposed to arise at the frequency $f_0$. These minor problems are resolved if we increase the thickness $t$, i.e., enlarging the nanoinsulator shields and therefore better confining the displacement current flows in the nanocircuit. This is reported in Fig. 28-30 for the case of $t = \lambda_0 / 150$. It is evident in this case that the features of the parallel interconnections between a nanoinductor and a nanocapacitor are all present in the plots: constant current across the elements with opposite phase between the two nanoparticles, same optical voltage drop across them both in amplitude and in phase and resonant peak at the design frequency $f_0$, confirming once again our heuristic analogy and the theoretical and numerical results of the previous sections for this complex 3D scenario.

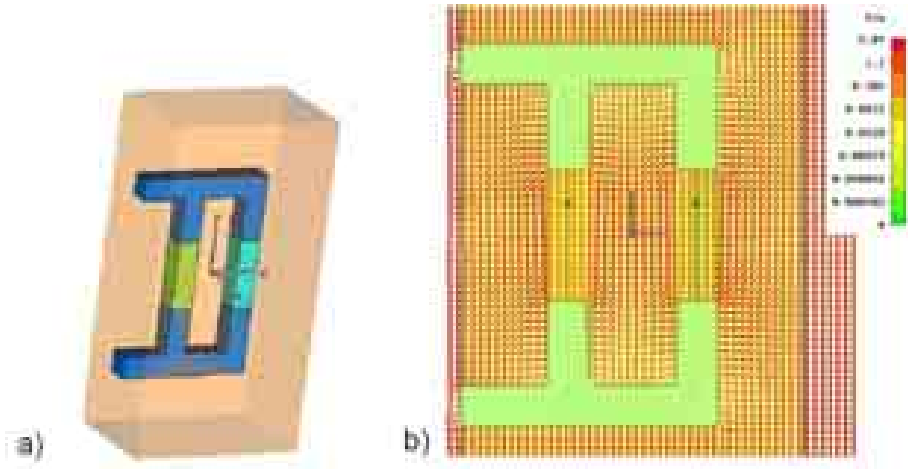

**Fig. 28.** (Color online) Similar to Fig. 25, but with larger thickness for the insulator region.

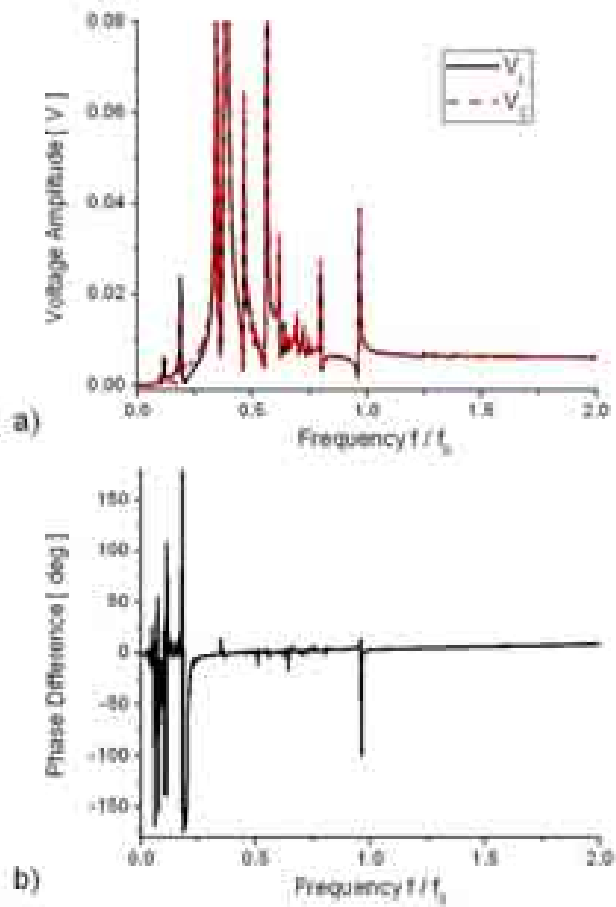

**Fig. 29.** (Color online) Same as Fig. 26, except with larger thickness for the insulator region.

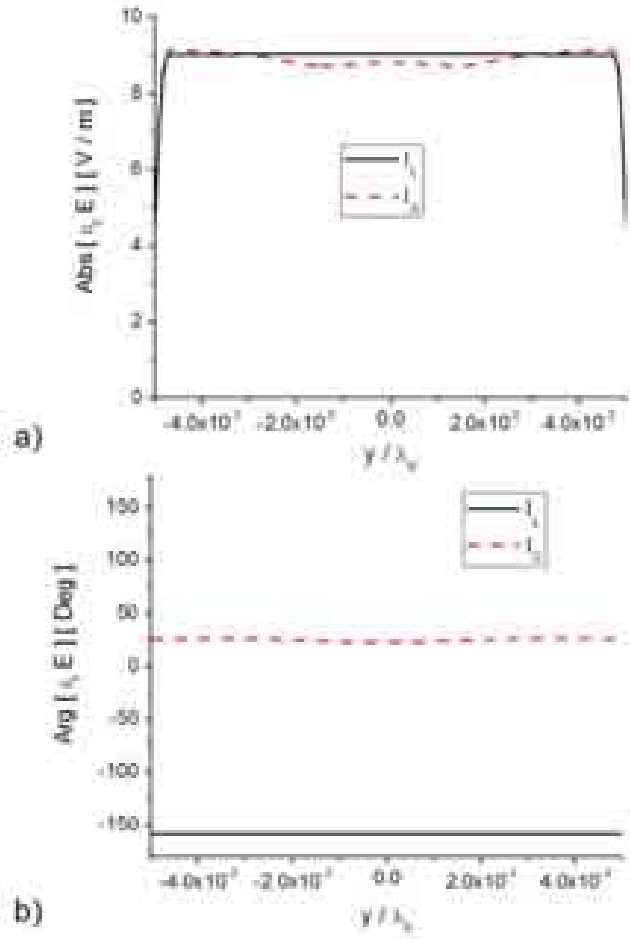

**Fig. 30.** (Color online) Same as Fig. 27, except with larger thickness for the insulator region.

It is worth underlining how the other peaks at lower frequencies in Fig. 27a, and even more pronounced in Fig. 30a, are clearly due to plasmonic resonances of the whole structure. It should be recalled that at frequencies below $f_0$, due to the necessary frequency dispersion of the ENZ nanoinsulators, such materials may have a negative permittivity, which may be characterized by a complex plasmonic response, depending on the geometry and electromagnetic properties of the whole nanocircuit. In our operation we are not interested in this frequency regime, since the nanocircuit can be designed for a desired frequency range above certain frequency $f_0$. It is also interesting to notice, however, that our circuit analogy applies over a relatively wide range of frequencies, despite the complexity of the scenario and the material dispersion and losses. These results may open novel possibilities in the realization of optical nanocircuits with filtering and guiding properties.

## VI. Unit Nanomodule of Lumped Optical Nanoelement

From the various simulations presented in the previous sections, we note that ENZ and EVL materials can play important roles as optical nanoinsulators and nanoconnectors for lumped nanoelements. This may lead us to the possibility of envisioning a "unit nanomodule" for lumped optical nanoelements. The deep subwavelength-size nanoparticle made of a plasmonic or non-plasmonic material can be insulated by thin layers of an ENZ material around its sides. The "top" and "bottom" ends of this insulated nanostructure may then be covered by thin layers of an EVL material. These EVL-covered "ends" may act as the "connecting points" for such an insulated nanoelement. Such a structure may then play the role of a unit nanomodule as a building block for a more complex optical nanocicruit. In order to assess the behavior of such a module, we report here some numerical simulations performed using the commercially available finite element method (FEM) software COMSOL Multiphysics®. For the sake of simplicity, we consider the geometry of this module to be two-dimensional (2D), i.e., the structure is assumed to be uniform along the axis normal to the plane of the paper. Since this module is assumed to be of deep sub-wavelength size, the FEM simulation is done under the "quasi-static" mode, similar to the study of the conventional low frequency (e.g., RF) electronic circuits. In this simulation, this unit module is placed between two perfectly electric conducting (PEC) parallel plates with an applied 1 volt potential difference between the two plates, so that a voltage drop is imposed on the module.

Although such a hypothetical feeding mechanism is not experimentally feasible, it does provide a mathematically easier configuration for FEM stimulation of optical field distribution of the nanomodule. Moreover, since the module is highly subwavelength and in a realistic excitation by a plane wave it will be effectively immersed in a locally uniform field, this feeding mechanism may provide reasonable assessment of the optical field distribution in and around this nanomodule, consistent with the full-wave (and more realistic) simulations reported in the previous section. An example is shown in Fig. 31, where the potential distributions inside and in the vicinity of the nanomodule are shown. The main material of the block is made of plasmonic material with $\varepsilon = -2\varepsilon_0$, acting as a nanoinductor. The color scheme represents the optical potential distribution, and the arrows show the direction (not the intensity) of displacement current. First we realize that the potential drop across this element has effectively opposite "phase" with the respect to the applied potential. This is due to the fact that the main material forming this nanoelement is an ENG, and thus this element acts as a nanoinductor. Moreover, we note that the displacement current flux leakage from the sides of this nanomodule is very low, and inside the nanoelement the displacement current is almost uniform and parallel with the side walls, indicating that very little flux leakage goes through the walls. An integral of the displacement vector over one of the side walls of this nanoelement (per unit length into the direction of normal to the paper) reveals that the flux leakage is around $2.33 \times 10^{-15}$ (C/m) while flux through one of the end point is about $2.52 \times 10^{-12}$ (C/m) – about three orders of magnitude difference – , which confirms the confinement of the displacement current inside the nanoelement. Since each end of the nanomodule is an equipotential surface, an optical "voltage drop" can be defined across this nanostructure.

Such a module may therefore have features that are mainly determined by the geometry and the constituent material properties, and they are essentially unaffected by the outside changes and relative orientations. In other words, such an optical nanoelement may have "modularized" functions, such as acting as a lumped impedance at optical frequencies.

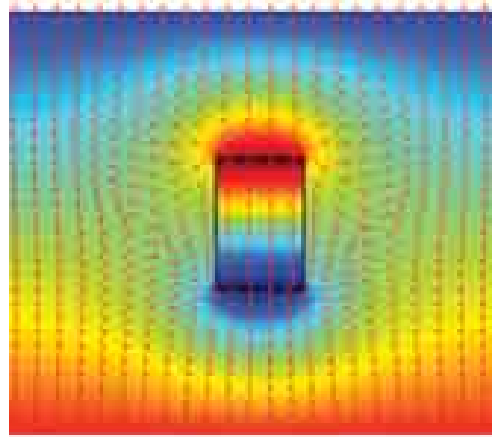

*(a)*

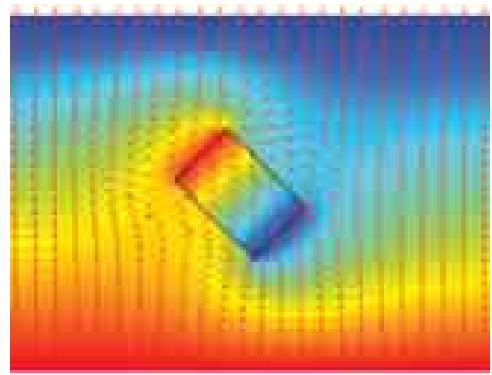

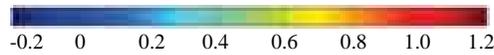

*(b)*

**Fig. 31.** Two-dimensional (2D) finite element method (FEM) "quasi-static" simulation of a unit nanomodule formed by a nanostructure made of a plasmonic or non-plasmonic material, insulated by very thin ENZ layers at the sides, and covered by very thin EVL layers on the top and bottom ends. (a) The case of nanoinductor module (with $\varepsilon_{ENG} = -2\varepsilon_o$). (b) the same as (a) but tilted. The color bar at the bottom is for both (a) and (b). Here the color shows the optical potential distributions, and the arrows show the direction of the displacement current.

Fig. 31*b* shows the response of the element when it is tilted so that the end cross sections are not necessarily parallel with the PEC plates. The displacement current inside the element still flows parallel to the side walls. This is a direct demonstration that the property of the element is intrinsic to the element and almost unperturbed by this tilt. To further highlight the modularity of these nanoelements, we have also examined a series combination of two of these building blocks (similar to what were discussed in the previous sections, but in the present FEM simulation environment.) Fig. 32 shows a series LC combination of two nanomodules, one with ENG ($\varepsilon_{ENG} = -2\varepsilon_o$) and the other with a conventional positive-epsilon materials ($\varepsilon_{DPS} = 2\varepsilon_o$). This effectively provides the resonance condition for this series LC combination and therefore, as expected, the voltage drops across each of these nanomodules have the same magnitude, but 180° out of phase. This is clearly evident in this figure by the same color at the two ends of the series LC. We also note that a considerable amount of displacement current goes through the two elements with almost no flux leakage from the side walls. Since $\varepsilon_1$ and $\varepsilon_2$ are of different sign, the electric field flips its direction when going across the interface, and therefore the line integral from one end of the series LC to the other end yields zero potential drop. FEM simulations have also been performed for the parallel combination of these two nanomodules, and the results (not shown here) support the expected functionalities for such a parallel interconnection.

The modularity of this nanoelement can be used to develop more complex optical nanocircuits in which the mathematical tools and machinery of circuit theory, such as Kirchhoff current and voltage laws, can be utilized at optical wavelengths, as discussed in previous sections. Fig. 33*a* shows a more complex nanocircuit formed by five nanomodules (four nanocapacitors and one nanoinductor), mimicking the function of the circuit shown in Fig. 33b. For the sake of simplicity in FEM simulation here, a 2D scenario has been considered. In this 2D configuration, the relative permittivities of different modules are shown with the values indicated by the color scale bar at the bottom, while the white segment represents EVL layers. The result of the FEM quasi-static simulation of this 2D configuration of five modules is shown in Fig. 33c, where the color scheme represents the optical potential, and the arrows show the direction of optical displacement current. The results are completely consistent with what is expected from the circuit shown in Fig. 33b, following the nanocircuit analogy described in the previous sections. Therefore, this example demonstrates that one may design optical lumped nanocircuits by arranging various optical nanomodules next to each other and form a tapestry of these nanostructures, providing circuit functionalities at optical frequencies.

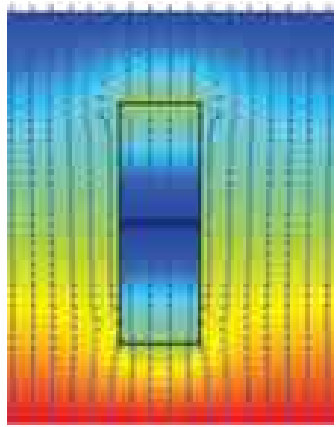

**Fig. 32.** Two-dimensional (2D) finite element method (FEM) "quasi-static" simulation of a series combination of two nanomodules, one being a nanoinductor (i.e., the upper element with $\varepsilon_{ENG} = -2\varepsilon_o$) and the other being a nanocapacitor (i.e., the bottom element with $\varepsilon_{DPS} = 2\varepsilon_o$). Each nanoelement is insulated by very thin ENZ layers in the sides and covered on the top and bottom ends by very thin EVL layers. Here the color scheme shows the optical potential distribution (same color scale as that in Fig. 31), and the arrows shows the direction (not amplitude) of the optical electric field. The 180º phase difference between the two potential differences across these two nanomodules (as expected in a series LC combination) can be clearly seen.

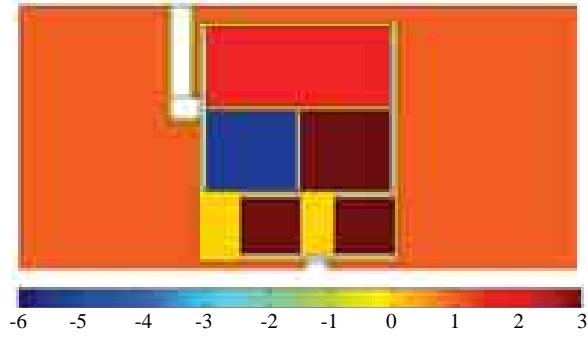

*(a)*

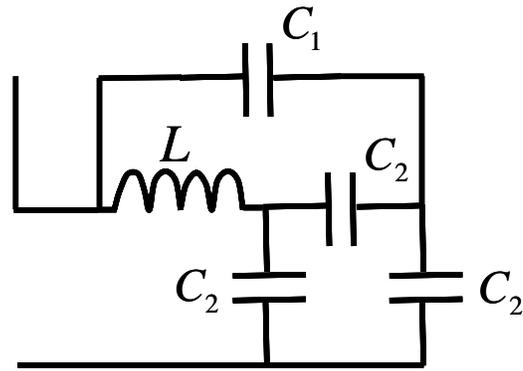

*(b)*

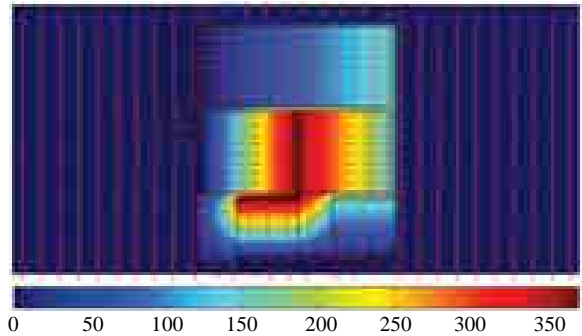

*(c)*

**Fig. 33.** (a) An optical nanocircuit formed by five nanomodules (four nanocapacitors and one nanoinductor), mimicking the function of the circuit shown in (b). Here a 2D configuration is considered. The value of the permittivity for each nanomodule is shown in the color scale in (a). The white region represents a material with a high permittivity (EVL). (c) Two-dimensional (2D) finite element method (FEM) "quasi-static" simulation of optical nanocircuit in (a). Here the color scheme shows the optical potential distributions, and the arrows shows the direction (not the amplitude) of displacement current in each nanomodule. We note how high the value of optical potential reaches in some of the nodes of this nanocircuit, due to the LC resonance.

-63-

# VII. Conclusions

In this work we have extended the concepts and formalism developed in [1], and studied under which conditions it is possible to characterize complex arrangements of (plasmonic and non-plasmonic) optical nanocircuit elements using the circuit theory. Different 2D and 3D circuit configurations that may be potentially interesting at optical and infrared frequencies have been investigated. We have developed accurate circuit models at optical wavelengths to characterize the equivalent impedance of the envisioned nanocapacitors and nanoinductors. It has also been shown that the induced displacement current may leak out of the subwavelength nanocircuit elements, causing strong coupling between the nanoelements and the neighboring region. To circumvent this problem, we have introduced the concept of optical nanoinsulators for the displacement current. It has been shown that by using ENZ materials it is possible to strongly reduce the unwanted displacement current leakage and confine the displacement current inside the nanocircuit. We have confirmed, both analytically and numerically, that nanocircuit elements insulated with ENZ materials may be accurately characterized using standard circuit theory concepts at optical frequencies, and in particular they may indeed be characterized by an equivalent impedance for nanocircuit elements. We have further explained how to apply the proposed circuit concepts in a scenario with realistic optical voltage sources. We have also studied how to ensure a good connection between the envisioned lumped nanoelements using EVL materials in the optical domain, and how this may avoid geometrical/polaritonic resonances at the junctions between the materials. This has led us to consider unit nanomodules for lumped nanocircuit elements, which may be regarded as building blocks for more complex nanocircuits at optical wavelengths. Finally, we have fully confirmed our predictions in 3D numerical simulations considering feeding models, frequency dispersion and presence of material losses. The new optical nanoinsulator and nanoconnector concepts, together with the results of [1], effectively show how to exploit and control the polaritonic resonances between subwavelength plasmonic and nonplasmonic elements, and they may establish the road map for designing complex nanocircuit arrangements at IR and optical frequencies that may be easily described using the well-known concepts of circuit theory, but at optical frequencies.

## Acknowledgments

This work is supported in part by the U.S. Air Force Office of Scientific Research (AFOSR) grant number FA9550-05-1-0442. Mário Silveirinha has been partially supported by a fellowship from "Fundação para a Ciência e a Tecnologia" during his stay at the University of Pennsylvania.

## Appendix A

Here we demonstrate how the complex electromagnetic problem under study in Section III.A (Fig. 2b) may be analytically solved in the $\varepsilon_{shield} = 0$ limit. To this end, we will use the theoretical formalism developed by us in [21]. To begin with, we reformulate the problem as a scattering problem. Due to the 2D-geometry, the magnetic field is of the form $\mathbf{H} = H_z \hat{\mathbf{u}}_z$, and the electric field may be derived from $H_z$: $\mathbf{E} = 1/(-i\omega\varepsilon)\nabla H_z \times \hat{\mathbf{u}}_z$. As referred in section II, the magnetic field radiated by the line sources is of the form (with $I_m = 1$, so that the induced electromotive force is 1[V] in the quasi-static limit):

$$\psi^{inc} \equiv H_z^{inc} = i\omega\varepsilon_0 \left( \Phi_0 \left( \mathbf{r} - \mathbf{r}'_+ \right) - \Phi_0 \left( \mathbf{r} - \mathbf{r}'_- \right) \right) \tag{A1}$$

where $\mathbf{r}'_+ = \left( R_s^+, 0, 0 \right)$ and $\mathbf{r}'_- = \left( R_s^-, 0, 0 \right)$. The total magnetic field $H_z$ all over the space is the sum of the incident field $\psi^{inc}$ and of the scattered field, which satisfies the usual radiation conditions. The total field satisfies:

$$\nabla \cdot \frac{1}{\varepsilon(\mathbf{r})} \nabla H_z + k_0^2 \mu_0 H_z = 0 \tag{A2}$$

As pointed out in [21], in the $\varepsilon_{shield} = 0$ limit the magnetic field is necessarily constant inside each connected ENZ-shield because otherwise the electric field inside the nanoinsulators would become arbitrarily large, and this may be proved impossible for energy considerations. Hence, we can write that:

$$H_z = H_z^{int,1}, \qquad R_1' < r < R_1 \tag{A3a}$$
$$H_z = H_z^{int,2}, \qquad R_2 < r < R_2' \tag{A3b}$$

where $H_z^{int,1}$ and $H_z^{int,2}$ are the (unknown) constant magnetic fields inside the interior shield (defined by $R_1' < r < R_1$) and exterior shield (defined by $R_2 < r < R_2'$), respectively. This result has an immediate and very important consequence: the electromagnetic fields inside the shielded ring may be written uniquely in terms of $H_z^{int,1}$ and $H_z^{int,2}$, independently of the specific source configuration (of course, the specific values of $H_z^{int,1}$ and $H_z^{int,2}$ depend indirectly on the source properties, as detailed ahead). Indeed, the magnetic field inside the ring, $R_1 < r < R_2$, is the solution of (A2) subject to the Dirichlet boundary conditions $H_z = H_z^{int,1}$ at $r = R_1$ and $H_z = H_z^{int,2}$ at $r = R_2$. Hence, in the $\varepsilon_{shield} = 0$ limit, the distribution of fields inside the shielded ring is completely independent of the distribution of the fields outside. This clearly evidences that the proposed nanoinsulators are able to effectively isolate the regions of the space that they delimit from other disturbances or field fluctuations, forcing the displacement current to circulate within the circuit region. Assuming that the permittivity of the ring is uniform, it can be easily proved that the solution of the previously mentioned Dirichlet problem is:

$$H_z = H_z^{int,1}\psi_1 + H_z^{int,2}\psi_2, \qquad R_1 < r < R_2 \tag{A4a}$$

$$\psi_1 = \frac{J_0(k_m R_2)Y_0(k_m r) - Y_0(k_m R_2)J_0(k_m r)}{J_0(k_m R_2)Y_0(k_m R_1) - Y_0(k_m R_2)J_0(k_m R_1)} \tag{A4b}$$

$$\psi_2 = \frac{J_0(k_m R_1)Y_0(k_m r) - Y_0(k_m R_1)J_0(k_m r)}{J_0(k_m R_1)Y_0(k_m R_2) - Y_0(k_m R_1)J_0(k_m R_2)} \tag{A4c}$$

where $k_m = \omega\sqrt{\varepsilon\mu_0}$ is the wave number inside the ring, and $J_n$ and $Y_n$ are the Bessel functions of 1$^{st}$ kind and order $n$. Again, it is worth underlining that the above formula is completely independent of the source position or configuration. Also, note that the magnetic field inside the shielded ring is independent of $\varphi$, since in the $\varepsilon_{shield} = 0$ limit the leakage is completely blocked. It is also interesting to note that the form of these results is independent on the thickness of the shields: as long as their permittivity is sufficiently close to zero any thickness of the two shields would support a solution for the internal problem given by (A4).

Similarly, as demonstrated in [21], outside the shielded ring the fields can also be written in terms of $H_z^{int,1}$ and $H_z^{int,2}$. Indeed, due to the linearity of the problem and using the superposition principle, the following equation holds in the free-space regions [21]:

$$H_z = \psi_{PMC} + H_z^{int,1}\psi_{s,1}^1, \qquad r < R_1' \tag{A5a}$$

$$H_z = \psi_{PMC} + H_z^{int,2}\psi_{s,2}^1, \qquad r > R_2' \tag{A5b}$$

where $\psi_{PMC}$ is the total magnetic field when the ENZ-shields are replaced by fictitious perfect magnetic conducting (PMC) materials (and the source configuration is unchanged), and $\psi_{s,2}^1$ is the radiating solution of (A2) subject to the boundary condition $\psi_{s,2}^1 = 1$ at $r = R_2'$ (the line sources are removed), and $\psi_{s,1}^1$ is defined analogously. It is straightforward to prove that:

$$\psi_{s,1}^1 = \frac{J_0(k_0 r)}{J_0(k_0 R_1')}, \qquad r < R_1' \tag{A6a}$$

$$\psi_{s,2}^1 = \frac{H_0^{(1)}(k_0 r)}{H_0^{(1)}(k_0 R_2')}, \qquad r > R_2' \tag{A6b}$$

On the other hand, consistently with its definition, the field $\psi_{PMC}$ can be written as:

$$\psi_{PMC} = \psi^{inc} + \begin{cases} c_0 J_0(k_0 r) + 2\sum_{n=1}^{\infty} c_n J_n(k_0 r)\cos(n\varphi), & r < R_1' \\ d_0 H_0^{(1)}(k_0 r) + 2\sum_{n=1}^{\infty} d_n H_n^{(1)}(k_0 r)\cos(n\varphi), & r > R_2' \\ 0 \quad \text{elsewhere} \end{cases} \tag{A7}$$

where $\psi^{inc}$ is given by (A1), and $c_n$, $d_n$, $n=0,1,2,\ldots$ are unknowns that can be determined by imposing that $\psi_{PMC} = 0$ at $r = R_1'$ and $r = R_2'$. Using the "addition theorem" for the Hankel function [23]:

$$H_0^{(1)}(k|\mathbf{r}-\mathbf{r}'|) = H_0^{(1)}(k r) J_0(k r') + 2\sum_{n=1}^{\infty} H_n^{(1)}(k r) J_n(k r')\cos(n\theta), \quad r > r' \tag{A8}$$

where $\theta$ is the angle defined by the vectors $\mathbf{r}$ and $\mathbf{r}'$, one can find that:

$$c_n = \frac{-\dfrac{\omega\varepsilon_0}{4} H_n^{(1)}(k_0 R_1') J_n(k_0 R_s^-)}{J_n(k_0 R_1')}, \quad d_n = \frac{+\dfrac{\omega\varepsilon_0}{4} H_n^{(1)}(k_0 R_s^+) J_n(k_0 R_2')}{H_n^{(1)}(k_0 R_2')} \tag{A9}$$

Equations (A3), (A4) and (A5) provide the exact solution of the electromagnetic problem in all space in terms of the unknowns $H_z^{int,1}$ and $H_z^{int,2}$. Using (A6), (A7), and (A9), the remaining parameters may be evaluated in closed analytical form. In order to obtain $H_z^{\text{int},1}$ and $H_z^{\text{int},2}$, we use the same procedure as in our previous work [21]. Namely, we apply Faraday's law to the boundary of each of the ENZ-materials. For example, for the ENZ-ring defined by $R_1' < r < R_1$, we obtain that:

$$\oint_{r=R_1} \mathbf{E}\cdot\mathbf{dl} - \oint_{r=R_1'} \mathbf{E}\cdot\mathbf{dl} = +i\omega\mu_0 H_z^{\text{int},1} A_{p,1} \tag{A10}$$

where $dl$ is the element of arc, $A_{p,1} = \pi\left(R_1^2 - R_1'^2\right)$ is the area of the interior ENZ-shield, and $A_{p,2}$ is defined analogously. A similar formula may be obtained for the exterior ENZ-ring. Using the continuity of the tangential component of the electric field at the interfaces, and (A4), (A5) and the formula $\mathbf{E} = (1/-i\omega\varepsilon)\nabla H_z \times \hat{\mathbf{u}}_z$, it is found after tedious but straightforward calculations that $H_z^{\text{int},1}$ and $H_z^{\text{int},2}$ verify the following linear system,

$$\left(k_0^2 A_{p,1} + \frac{2\pi R_1}{\varepsilon_r}\frac{\partial \psi_1}{\partial r}\bigg|_{r=R_1} - 2\pi R_1'\frac{\partial \psi_{s,1}^1}{\partial r}\bigg|_{r=R_1'}\right) H_z^{\text{int},1} + \left(\frac{2\pi R_1}{\varepsilon_r}\frac{\partial \psi_2}{\partial r}\bigg|_{r=R_1}\right) H_z^{\text{int},2} = \\ 2\pi R_1'\frac{\partial}{\partial r}\left(\frac{\omega\varepsilon_0}{4} J_0\left(k_0 R_s^-\right) H_0^{(1)}\left(k_0 r\right) + c_0 J_0\left(k_0 r\right)\right)\bigg|_{r=R_1'} \tag{A11a}$$

$$\left(-\frac{2\pi R_2}{\varepsilon_r}\frac{\partial \psi_1}{\partial r}\bigg|_{r=R_2}\right) H_z^{\text{int},1} + \left(-\frac{2\pi R_2}{\varepsilon_r}\frac{\partial \psi_2}{\partial r}\bigg|_{r=R_2} + k_0^2 A_{p,2} + 2\pi R_2'\frac{\partial \psi_{s,2}^1}{\partial r}\bigg|_{r=R_2'}\right) H_z^{\text{int},2} = \\ -2\pi R_2'\frac{\partial}{\partial r}\left(-\frac{\omega\varepsilon_0}{4} H_0^{(1)}\left(k_0 R_s^+\right) J_0\left(k_0 r\right) + d_0 H_0^{(1)}\left(k_0 r\right)\right)\bigg|_{r=R_2'} \tag{A11b}$$

where $\varepsilon_r$ is the relative permittivity of the ring. The solution of the above system yields the desired $H_z^{\text{int},1}$ and $H_z^{\text{int},2}$, and this formally solves the problem under-study in closed analytical form. Since the formulas for $H_z^{\text{int},1}$ and $H_z^{\text{int},2}$ are rather cumbersome, it is instructive to use the derived results to obtain a quasi-static solution of the problem, valid when the dimensions of the ring are subwavelength. In the quasi-static limit, (A11) simplifies to (retaining only the dominant powers of $k_0 R$):

$$\begin{pmatrix} -\dfrac{2\pi}{\ln(R_2/R_1)}\dfrac{1}{\varepsilon_r} & \dfrac{2\pi}{\ln(R_2/R_1)}\dfrac{1}{\varepsilon_r} \\ \dfrac{2\pi}{\ln(R_2/R_1)}\dfrac{1}{\varepsilon_r} & -\dfrac{2\pi}{\ln(R_2/R_1)}\dfrac{1}{\varepsilon_r} - \dfrac{H_1^{(1)}(k_0 R_2')}{H_0^{(1)}(k_0 R_2')} 2\pi R_2' k_0 \end{pmatrix} \begin{pmatrix} H_z^{\text{int},1} \\ H_z^{\text{int},2} \end{pmatrix} = \begin{pmatrix} i\omega\varepsilon_0 \dfrac{J_0(k_0 R_s^-)}{J_0(k_0 R_1')} \\ -i\omega\varepsilon_0 \dfrac{H_0^{(1)}(k_0 R_s^+)}{H_0^{(1)}(k_0 R_2')} \end{pmatrix}$$

(A12)

The solution of the system in the asymptotic limit $k_0 R \to 0$ is:

$$\begin{pmatrix} H_z^{\text{int},1} \\ H_z^{\text{int},2} \end{pmatrix} \approx (-i\omega\varepsilon_0) \begin{pmatrix} \dfrac{\ln(R_s^+/R_2')}{2\pi} + \dfrac{\ln(R_2/R_1)\varepsilon_r}{2\pi} \\ \dfrac{\ln(R_s^+/R_2')}{2\pi} \end{pmatrix} \quad (A13)$$

On the other hand, using (A4) and $\mathbf{E} = (1/-i\omega\varepsilon)\nabla H_z \times \hat{\mathbf{u}}_z$, and letting $k_0 R \to 0$, one can find that:

$$\mathbf{E} \approx \dfrac{1}{r\varepsilon_r} \dfrac{1}{\ln\left(\dfrac{R_2}{R_1}\right)} \dfrac{-1}{i\omega\varepsilon_0} \left( H_z^{\text{int},1} - H_z^{\text{int},2} \right) \hat{\mathbf{u}}_\varphi \quad , \quad R_1 < r < R_2 \quad (A14)$$

Hence, substituting (A13) into (A14), one finds that:

$$\mathbf{E} \approx \dfrac{1}{2\pi r} \hat{\mathbf{u}}_\varphi, \quad R_1 < r < R_2 \qquad \text{(quasi-static limit)} \quad (A15)$$

Thus, consistently with our expectations, in the quasi-static limit the electric field inside the ring only has an azimuthal component and is such that the induced electromotive force is $V = 1\,[\text{V}]$. Using the above formula, one can easily find that the electric flux (p.u.l) inside the ring is:

$$\phi_e \approx \frac{1}{2\pi} \ln\left(\frac{R_2}{R_1}\right) \varepsilon \approx \frac{1}{2\pi} \frac{R_2 - R_1}{R_{med}} \varepsilon \tag{A16}$$

being the second identity valid if $\frac{R_2 - R_1}{R_{med}} \ll 1$. Therefore the electrical reluctance is given by:

$$\Re_e \equiv \frac{V}{\phi_e} \approx \frac{2\pi}{\varepsilon \ln\left(\frac{R_2}{R_1}\right)} \approx \frac{2\pi R_{med}}{\varepsilon (R_2 - R_1)} \tag{A17}$$

One can recognize that the above result, obtained directly from the exact solution of the problem, is coincident with the formulas derived in section II, and thus supports our circuit analogy.

## Appendix B

We show here formally that the ENZ nanoinsulators, in the limit of $\varepsilon_{shield} = 0$, prevent the excitation of SPPs for the parallel geometry of Fig. 10. Consider, in fact, that the permittivity of the shielded region in this case is of the form $\varepsilon = \varepsilon(r)$, i.e., the permittivity only depends on the radial coordinate. Note that the two-layer structure that we characterized before (in which the permittivity $\varepsilon = \varepsilon(r)$ only assumes two different values, $\varepsilon_{in}$ and $\varepsilon_{out}$) is a particular case of this much more general configuration. As in Appendix A, it is possible to solve the electromagnetic problem under study in closed analytical form in the limit of $\varepsilon_{shield} = 0$. In particular, using the same arguments as in Appendix A, it is clear that the magnetic field $H_z$ inside the shielded region is given by the solution of (A2) (with $\varepsilon = \varepsilon(r)$) subject to the Dirichlet boundary conditions $H_z = H_z^{int,1}$ at $r = R_1$ and $H_z = H_z^{int,2}$ at $r = R_2$. This simple observation has a very important consequence: indeed, since both the shielded domain and the permittivity $\varepsilon = \varepsilon(r)$ are invariant to rotations, it is clear that the solution of the mentioned Dirichelet problem is invariant as well, i.e., $H_z = H_z(r)$ in $R_1 < R < R_2$, and in particular the electric field only has an azimuthal component in the same region, $\mathbf{E} = E_\varphi(r)\hat{\mathbf{u}}_\varphi$. In other words, in the $\varepsilon_{shield} = 0$ limit and for $\varepsilon = \varepsilon(r)$, the field distribution inside the shielded region $R_1 < R < R_2$ is invariant to rotations, *independently* of the source configuration or of the specific source position. Hence, it follows that in the $\varepsilon_{shield} = 0$ limit it is impossible to excite SPPs in the interface between different material layers, even when these layers are not interleaved with an ENZ-nanoinsulator. These facts simplify the parallel configuration of the proposed nanoinsulators and they may help confining the displacement current inside the circuit path and reduce the coupling between the subwavelength rings.

As mentioned above, by proceeding as in Appendix A it is possible to solve the electromagnetic problem under study in closed analytical form in the limit of $\varepsilon_{shield} = 0$. However, it is more informative to derive an approximate solution valid in the quasi-static limit, as we do in the following. We know that in this limit the induced electromotive force inside the shielded region is approximately $V = 1$ [V] (for the considered source configuration). Also, as pointed out previously, in the $\varepsilon_{shield} = 0$ limit and for $\varepsilon = \varepsilon(r)$ the electric field is exactly of the form $\mathbf{E} = E_\varphi(r)\hat{\mathbf{u}}_\varphi$. But these two elementary facts imply that in the quasi-static limit the electric field necessarily verifies (A15). Hence, it follows from the definition that the induced total flux (p.u.l) is given by:

$$\phi_e = \int D \cdot \mathbf{ds} = \int_{R_1}^{R_2} \varepsilon(r) E_\varphi(r) dr = \frac{1}{2\pi} \int_{R_1}^{R_2} \frac{\varepsilon(r)}{r} dr \qquad (B1)$$

Hence, the equivalent reluctance $\mathfrak{R}_e = V/\phi_e$ is such that:

$$\frac{1}{\mathfrak{R}_{e,eq}} \approx \frac{1}{2\pi} \int_{R_1}^{R_2} \frac{\varepsilon(r)}{r} dr \quad \text{(p.u.l)} \qquad (B2)$$

Very interestingly, the above formula demonstrates that the equivalent impedance is the parallel combination of the impedances of each (infinitesimal/uniform) section of the ring. Note that this result is completely consistent with (6), which applies when the shielded region consists of two uniform rings with permittivity $\varepsilon_{in}$ and $\varepsilon_{out}$. This further supports our theory and the possibility of characterizing these nanostructures using circuit theory. In the general case of parallel combinations of nanocircuit elements of more arbitrary shape, the presence of a further shield at the interface between parallel elements may help preventing any undesired coupling, current exchange or local polariton excitation, even though, as we have shown in this Appendix, in this specific configuration such extra shield is unnecessary.

## Appendix C

Here we formally derive the quasi-static analytical solution of the problem highlighted in Section IIIC of the series interconnection of Fig. 11 in the limit of $\varepsilon_{shield} = 0$. To this end, we admit that the permittivity of the shielded ring is of the form $\varepsilon = \varepsilon(\varphi)$, i.e., the permittivity depends uniquely on the azimuthal angle (the geometry depicted in the inset of Fig. 11 corresponds to the particular case in which $\varepsilon(\varphi)$ only assumes two values: $\varepsilon_1$ and $\varepsilon_2$). As explained in Appendix A, when $\varepsilon_{shield} = 0$ the exact solution for $H_z$ inside the ring is the solution of (A2) (with $\varepsilon = \varepsilon(\varphi)$) subject to the Dirichlet boundary conditions $H_z = H_z^{int,1}$ at $r = R_1$ and $H_z = H_z^{int,2}$ at $r = R_2$. In general, it is not possible to obtain the solution of this problem in closed analytical form following the steps of Appendix A. However, in the quasi-static limit we can neglect in first approximation the second parcel in the left-hand side of (A2). Under this approximation the solution of (A2), subject to the indicated boundary conditions, is:

$$H_z \approx \frac{H_z^{int,1} \ln\left(\frac{r}{R_2}\right) - H_z^{int,2} \ln\left(\frac{r}{R_1}\right)}{\ln\left(\frac{R_1}{R_2}\right)} \tag{C1}$$

Using $\mathbf{E} = (1/-i\omega\varepsilon)\nabla H_z \times \hat{\mathbf{u}}_z$, it is found that the corresponding electric field verifies:

$$\mathbf{E} \approx \frac{1}{r} \frac{1}{\ln\left(\frac{R_2}{R_1}\right)} \frac{1}{-i\omega\varepsilon(\varphi)} \left(H_z^{int,1} - H_z^{int,2}\right) \hat{\mathbf{u}}_\varphi \quad , \quad R_1 < r < R_2 \tag{C2}$$

Hence the induced electromotive force and the induced flux (p.u.l) are given by:

$$V \approx \frac{2\pi}{\ln\left(\frac{R_2}{R_1}\right)} \frac{1}{-i\omega\varepsilon_\parallel} \left(H_z^{int,1} - H_z^{int,2}\right), \qquad \frac{1}{\varepsilon_\parallel} = \frac{1}{2\pi} \int_0^{2\pi} \frac{1}{\varepsilon(\varphi)} d\varphi \tag{C3a}$$

$$\phi_e \approx \frac{1}{-i\omega\varepsilon(\varphi)} \left(H_z^{int,1} - H_z^{int,2}\right) \tag{C3b}$$

The above formulas show that the equivalent reluctance, $\mathfrak{R}_e = V/\phi_e$, of the non-uniform ring characterized by the permittivity $\varepsilon = \varepsilon(\varphi)$ is given by:

$$\mathfrak{R}_e = \frac{2\pi}{\varepsilon_\parallel \ln\left(\frac{R_2}{R_1}\right)} \approx \frac{2\pi R_{med}}{\varepsilon_\parallel (R_2 - R_1)} = \frac{R_{med}}{(R_2 - R_1)} \int_0^{2\pi} \frac{1}{\varepsilon(\varphi)} d\varphi \tag{C4}$$

-74-

In particular, we conclude that in the quasi-static limit the equivalent impedance is the series combination of the impedances of each (infinitesimal/uniform) section of the ring, consistently with formula (8) for a sectionally constant permittivity.